\documentclass[11pt,a4paper]{article}
\usepackage{jheppub}
\usepackage[utf8]{inputenc}
\usepackage{physics}
\usepackage{bbm}
\newcommand{\av}[1]{\ensuremath{\langle#1\rangle}}
\renewcommand{\exp}[1]{\,\mathrm{exp}\left[#1\right]}
\newcommand{\beq}{\begin{equation}\begin{aligned}}
\newcommand{\eeq}{\end{aligned}\end{equation}}
\newcommand{\bqa}{\begin{eqnarray}}
\newcommand{\eqa}{\end{eqnarray}}

\newcommand{\ir}{{\text{\tiny IR}}}
\newcommand{\uv}{{\text{\tiny UV}}}
\usepackage{xcolor}
\usepackage{soul}
\usepackage[export]{adjustbox}
\usepackage[final]{pdfpages}
\usepackage{subcaption}
\usepackage{pgfplots}
\usepackage{adjustbox}
\pgfplotsset{compat=1.9}
\pgfplotsset{
    tick label style={font=\footnotesize},
    label style={font=\footnotesize},
    legend style={font=\tiny},
    every axis plot/.append style={line width=0.5pt},
    }
\title{Stochastic inflation from quantum field theory and the parametric dependence of the effective noise amplitude}

\author[a]{Jens O. Andersen,}
\author[a,b]{Magdalena Eriksson,}
\author[b]{Anders Tranberg}
\affiliation[a]{Department of Physics, Faculty of Natural Sciences,  
Norwegian University of Science and Technology, H{\o}gskoleringen 5,
N-7491 Trondheim, Norway}
\affiliation[b]{Faculty of Science and Technology, University of Stavanger,
  4036 Stavanger, Norway}
\emailAdd{jens.andersen@ntnu.no}
\emailAdd{magdalena.eriksson@ntnu.no}
\emailAdd{anders.tranberg@uis.no}
\abstract{
The non-linear dynamics of long-wavelength cosmological fluctuations may be phrased in terms of an effective classical, but stochastic evolution equation. The stochastic noise represents short-wavelength modes that continually redshift into the long-wavelength domain. The effective evolution may be derived from first principles quantum field theory in an expanding background, through a sequence of approximations calling for additional scrutiny. We perform such an analysis, putting particular emphasis on the amplitude of the stochastic noise, which ultimately determines the cosmological correlations and provides a non-perturbative IR regulator to the dynamics. }
\begin{document}
\maketitle

\section{Introduction}
\label{sec:introduction}
The primordial density perturbations in the Universe are expected to have originated as vacuum fluctuations of a light, weakly-interacting scalar field $\phi$, amplified during a period of accelerating cosmological expansion known as inflation (see for instance \cite{Riotto:2002yw,LythLiddle}). However, it has long been known that when computing cosmological correlations in perturbative quantum field theory (QFT), the (near-)massless propagator leads to unphysical divergent and secular IR behaviour, which call for regularisation \cite{Tsamis:2005hd,vanderMeulen:2007ah}. 

One way to achieve this is through resumming infinite sets of perturbative diagrams, using resummation techniques well-known from Minkowski space computations, adapted to de Sitter space (dS). These include the large-$N$ expansion \cite{Serreau:2011fu,Serreau:2013psa}, truncations of the two-particle-irreducible (2PI) effective action formalism \cite{Arai:2011dd,LopezNacir:2013alw}, the non-perturbative renormalisation group (RG) \cite{Serreau:2013eoa}, and dressing of the Euclidean zero mode by summing an infinite class of diagrams
\cite{Rajaraman:2010xd,Beneke:2012kn,LopezNacir:2016gzi}. These all reveal that a dynamical mass is generated by non-linear interactions, regularising the correlation functions, even in the case of minimally coupled massless scalar fields. 

An alternative to the diagrammatic QFT analysis is stochastic inflation \cite{Starobinsky:1986fx,Starobinsky:1994bd}, where a separation of scales is introduced between the (far) super-horizon (IR) modes and the near- and sub-horizon (UV) modes. By integrating out the UV modes, an effective classical, but stochastic, IR theory arises, which retains much of the nonlinear, nonperturbative dynamics of the exact theory. The IR dynamics again generate a dynamical mass.
This is reminiscent of similar approaches to quantum thermal field theories of gauge and fermion fields \cite{bodeker}, where UV degrees of freedom source classical IR dynamics in the form of a stochastic noise. One important difference is that in stochastic inflation, the noise is the result of a sliding scale separation turning UV modes into IR modes. It therefore appears even in the non-interacting theory, and is hence unsuppressed by powers of the couplings. 
It also follows that the stochastic approach is specific to near-de Sitter geometries in that the Hubble horizon is used as a separation scale between the long and short wavelength modes, with modes leaving the horizon only for an accelerating expansion. 

In the simplest overdamped (slow-roll) limit, $\Ddot{\phi}\ll 3H\Dot{\phi}$, and in addition neglecting spatial gradients, 
the effective IR dynamics can be described by a Langevin equation, and field correlations be obtained via the corresponding Fokker-Planck equation. This is the framework most often applied to models of inflation, although the problem is still numerically tractable even when relaxing some of these assumptions \cite{Cable:2020dke}.

The stochastic approach has been shown to reproduce the correct IR behaviour of the full QFT statistical propagator to leading order in the coupling
\cite{Tsamis:2005hd,Finelli:2008zg,Garbrecht:2013coa,Garbrecht:2014dca,Onemli:2015pma} and was recently favourably compared at two-loop order in perturbation theory \cite{Kamenshchik:2021tjh}. Other authors have studied the evolution of the density matrix for the IR modes to recover a Fokker-Planck equation \cite{Burgess:2015ajz,Collins:2017haz}, and Ref. \cite{Baumgart:2019clc} argued the emergence of the stochastic formalism from a leading-log diagrammatic analysis at coincident points.  The conceptual and computational tractability offered by the stochastic approach has made it a popular method (see e.g. \cite{Karakaya:2017evp,Grain:2017dqa} for example applications), and in recent years there has been an upsurge in activity aiming to extend it e.g. to beyond leading order in the coupling \cite{Collins:2017haz,Tokuda:2017fdh,Cohen:2021fzf}, the slow-roll approximation \cite{Pattison:2019hef,Cable:2020dke}, or to include derivative interactions \cite{Kitamoto:2018dek}. However, compared to its extensive use, studies of the embedding and range of validity of the stochastic approach in QFT remain scarce (see however \cite{Morikawa1990,Matarrese:2003ye}), which has motivated the present work. 

In this paper we will revisit the first-principles derivation at the level of the path integral, paying attention to the sequence of approximations required to reach the standard form of the effective dynamics. These include the coarse graining procedure, the scale separation specified by a so-called window function and the parametrisation of the free UV vacuum sourcing the horizon-crossing modes. In particular, the stochastic noise is often taken to have a particular, universal amplitude, which we will see only arises in some very specific parametric limits.

The article is organised as follows. In the following subsection \ref{sec:langevin}, we present the effective stochastic theory, and how it is applied to compute IR field correlators given an inflation model. Having described the objective of our discussion, we in section \ref{sec:QFTder} introduce the real-time QFT formalism for the problem at hand, and show how an effective IR theory arises (sections \ref{sec:coarsegraining} and \ref{sec:IReffective}) and how to interpret it as a stochastic theory (section \ref{sec:IRstochastic}). We reconnect to the standard form of the stochastic noise resulting from the choice of a sharp window function applied to the field variables in momentum space \cite{Starobinsky:1994bd} (section \ref{sec:Starobinsky}). This allows us to identify the set of assumptions and approximations going into the derivation, and their validity. We investigate an alternative mode separation procedure at the level of the path integral in section \ref{sec:propagatordecomp}, and proceed to again derive the target theory, in the process identifying the approximations required. The procedure is illustrated and approximations checked through an example (Gaussian) window function in section \ref{sec:Testing}.
In section \ref{sec:noiseamplitude} we compute and compare the amplitude of the stochastic noise for a selection of window functions, considering both massless and massive UV modes, dS and slow-roll spacetimes, and test the validity of the central assumptions. Conclusions are gathered in section \ref{sec:conclusion}.

\subsection{Application of stochastic dynamics to inflationary perturbations }
\label{sec:langevin}

We first briefly review the stochastic inflationary formalism, and how a dynamical mass is generated. The dynamics of the coarse-grained IR field $\phi_\ir$ on super-Hubble scales is described 
by a Langevin equation
\beq
    \dot{\phi}_\ir(t)+\frac{V'(\phi_\ir)}{3H} = \xi_\phi(t), 
    \label{eq:Langevin1}
\eeq
where $H\equiv \dot{a}(t)/a(t)$ is the Hubble rate, 
\beq
    V(\phi)=\tfrac{1}{2}m^2\phi^2 + V_\text{I}(\phi),
\eeq
is the potential and $\xi_\phi$ is a Gaussian stochastic noise with the correlation
\beq
    \av{\xi_\phi(t)\xi_\phi(t')} = \frac{H^3}{4\pi^2}f(\epsilon_H,\epsilon_M,\dotso)\delta(t-t'),
    \label{StarobinskyNoiseCorrelator}
\eeq
corresponding to white, Markovian, noise statistics. Most commonly, the function $f$ is taken to be trivial, $f=1$, but we will allow for the noise amplitude to a priori depend on several
variables, including the slow-roll and mass parameters 
$\epsilon_H$ and $\epsilon_M$ defined as
\beq
\epsilon_H \equiv -\frac{\dot{H}}{H^2},\qquad \epsilon_M\equiv\frac{m^2}{3H^2}.
\label{eq:epsM}
\eeq 
The Langevin equation gives rise to a Fokker-Planck equation for the one-point probability distribution of the IR field 
\beq
    \partial_t P(t,\phi_\ir) = \left(\frac{V''(\phi_\ir)}{3H}+\frac{V'(\phi_\ir)}{3H}\partial_{\phi_\ir}+\frac{H^3f}{8\pi^2}\partial^2_{\phi_\ir}\right)P(t,\phi_\ir).
\eeq
Given an initial distribution at the beginning of inflation, the distribution at any later time then follows, and observables may be computed as
\beq
\langle\mathcal{O}(\phi_\ir(t))\rangle = \int \dd \phi_\ir P(t,\phi_\ir)\mathcal{O}(\phi_\ir).
\eeq 
Often, the late-time equilibrium solution is assumed to have been reached\footnote{Although this may not always be the case~\cite{Enqvist:2012,Hardwick:2017fjo}.}
\beq
    P_\text{eq}(\phi_\ir)\sim \exp{-\frac{8\pi^2V(\phi_\ir)}{3fH^4}}, \label{eq_PeqDistribution}
\eeq
and for instance for a free, massive theory, $V(\phi)=\tfrac{1}{2}m^2\phi^2$, the variance is given by
\beq
    \langle \phi^2_\ir\rangle= \frac{3fH^4}{8\pi^2 m^2}.
    \label{eq:massive}
\eeq
For $f=1$, this is indeed the finite part of the field correlator in the Bunch-Davies vacuum at leading order in the limit $m^2\ll H^2$. 
This can be seen directly from expanding the free semi-dS propagator in the same limit \cite{Prokopec:2003tm,Janssen:2007ht}, 
\beq
     G_F(y) 
     = \frac{H^2}{4\pi^2}\left(\frac{1}{y}-\frac{1}{2}\ln y+\frac{1}{2(\epsilon_M-\epsilon_H)}-1+\ln 2+\mathcal{O}(\epsilon_M,\epsilon_H)\right),
    \label{eq:massiveQFTprop}
\eeq
where $y$ is the dS invariant length scale, 
\beq
    y(x,x') 
    =4\sin^2\left(\tfrac{1}{2}l(x,x')H\right), 
\eeq
with $l(x,x')$ denoting the geodesic distance. Setting $\epsilon_H=0$ and inserting (\ref{eq:epsM}) into the leading finite term, 
we recover (\ref{eq:massive}).

However, when applying the stochastic formalism to a massless self-interacting field with
$V_\text{I}(\phi)=\frac{\lambda}{4}\phi^4$,
the late-time Fokker-Planck distribution straightforwardly leads to \cite{Starobinsky:1994bd}
\beq
    \lim_{t\rightarrow\infty}\av{\phi^2_\ir(t)} = \sqrt{\frac{3f}{2\pi^2}}\frac{\Gamma\left[\tfrac{3}{4}\right]}{\Gamma\left[\tfrac{1}{4}\right]}\frac{H^2}{\sqrt{\lambda}}. \label{eq_StarobVariance}
\eeq
One may then infer that a mass has been generated dynamically by comparing 
the massive dS result (\ref{eq:massive})
with \eqref{eq_StarobVariance}, 
i.e.,
\beq
    m^2_\text{dyn} = \frac{\sqrt{6}}{\sqrt{f}8\pi}\frac{\Gamma\left[\tfrac{1}{4}\right]}{\Gamma\left[\tfrac{3}{4}\right]}\sqrt{\lambda}H^2\simeq 0.288\frac{\sqrt{\lambda}}{\sqrt{f}}H^2.
\eeq
The effective mass squared is proportional to $H^2$ and importantly to $\sqrt{\lambda}$, rather than an integer power of the coupling. 
This non-analytic dependence suggests that the stochastic prescription amounts to a resummation of Feynman diagrams from all orders of perturbation theory, providing an IR regulator in an expanding background ($H\neq 0$). We see that the amplitude of the noise, including the function $f(\dotso)$, sets the scale for correlators and the dynamical mass. 

The original stochastic formalism was restricted to the slow-roll limit where $\ddot{\phi}_\ir$ is neglected, and where also $V^{\prime\prime}/H^2\ll 1$. However, one may generalise to include second-order derivatives in time
by introducing a "momentum noise" $\xi_\pi$ to accompany the new degree of freedom. The Langevin dynamics can then be written as two coupled equations for the IR field and its time derivative:
\beq 
    \dot{\phi}_\ir = \pi_\ir + \xi_\phi, \qquad \dot{\pi}_\ir+3H\pi_\ir+V'(\phi_\ir)=\xi_\pi,
    \label{eq:LangevinField_Momenta}
\eeq
alternatively,
\beq
    \ddot{\phi}_\ir+3H\dot{\phi}_\ir+V'(\phi_\ir)=3H\xi_\phi+\dot{\xi}_\phi+\xi_\pi \equiv \xi.
    \label{eq:twonoise}
\eeq
For later reference, we note that in the particular case, when $3H\xi_\phi\gg \dot{\xi}_\phi,\xi_\pi$ (in a sense to be discussed below), the normalisation (\ref{StarobinskyNoiseCorrelator}) implies
\beq
\av{\xi(t)\xi(t')} \rightarrow 9H^2\av{\xi_\phi(t)\xi_\phi(t')}= \frac{9H^5}{4\pi^2}f(\epsilon_H,\epsilon_M,\dotso)\delta(t-t').
\label{Starobxixi}
\eeq
In the following we will see that these results emerge also (in certain limits) in QFT. Starting from the closed-time-path (CTP) action, we will derive the Langevin equation using a suitable set of approximations. In particular we keep track of the normalisation of the noise, parametrised by a function $f(\dotso)$, which as we have seen appears prominently in the late-time Fokker-Planck distribution and hence the computed observables. The variance for the non-interacting dS vacuum corresponds to $f=1$ to leading order in the mass, and provides a natural benchmark for comparison.

\section{Stochastic dynamics from quantum field theory}
\label{sec:QFTder}

We first revisit the derivation of the Langevin evolution from a field theoretical perspective, paying special attention to the scale separation between the long- and short-wavelength degrees of freedom \cite{Starobinsky:1994bd,Moss:2016uix,Morikawa1990}.

\subsection{The closed-time-path formalism for inflation}
\label{sec:CTP}

We consider the real-time evolution of an interacting scalar field in an FRLW spacetime, 
whose line element is given by 
\beq
    \dd s^2=-\dd t^2+a^2(t)\dd\mathbf{x}^2,
\eeq
where $a(t)$ is the scale factor  and the slow-roll parameter $\epsilon_H$
parameterises the deviation from a constant expansion rate, corresponding to dS space where $a(t)\propto e^{Ht}$. 
Expectation values for operators are computed using the CTP formalism\footnote{Also known as the in-in or Schwinger-Keldysh formalism.} of QFT \cite{Chou84,Calzetta:1986ey}. In the CTP formalism, the time coordinate runs on a closed time path from $t_\text{in}$ to $t$ (the '+' branch) and back again (the '$-$' branch). The field $\phi$ and source $J$ are split up into path-ordered constituents $\phi^\pm$ and $J^\pm$, where for equal time $\phi^+(t)=\phi^-(t)$. The quantum correlators evaluated at time $t$ can be obtained by functional differentiation of the generating functional
\beq
    Z[\mathbbm{J}]=\int\mathcal{D}\Phi \exp{\frac{i}{2}\int_x\left(\Phi^T\mathbbm{G}^{-1}\Phi-2V_\text{I}(\Phi)+\mathbbm{J}^T\Phi\right)}, 
    \label{eq:genfungeneral}
\eeq
where $\int_x\equiv \int\dd^4x\sqrt{-g(x)}$ for brevity, $V_\text{I}(\Phi)=V_\text{I}(\phi^+)-V_\text{I}(\phi^-)$ is the interacting potential and
we define
\beq
    \Phi(x)\equiv \begin{bmatrix}
    \phi^+(x)\\\phi^-(x)
    \end{bmatrix} , \quad 
    \mathbbm{G}^{-1}(x,x') \equiv \begin{bmatrix}
    G^{-1}(x) & 0 \\ 0 & -G^{-1}(x)
    \end{bmatrix}\delta(x,x'),  \quad \mathbbm{J}(x)\equiv \begin{bmatrix}
    J^+(x)\\-J^-(x)
    \end{bmatrix}.
\eeq
The free inverse propagator is $G^{-1}(x)=\Box_x-m^2$ with $\Box_x\equiv \partial_\mu(\sqrt{-g(x)}g^{\mu\nu}\partial_\nu)/\sqrt{-g(x)}$. 
The matrix $\mathbbm{G}(x,x')$ consists of the CTP propagators with all four possible time orderings:
\beq
    \mathbbm{G}(x,x') = \begin{bmatrix}
    G^{++}(x,x') & G^{+-}(x,x') \\ G^{-+}(x,x') & G^{--}(x,x')
    \end{bmatrix} 
    = -i\begin{bmatrix}
    \av{T\phi^+(x)\phi^+(x')} & \av{\phi(x)^+\phi^-(x')}\\ \av{\phi^-(x')\phi^+(x)} & \av{\Bar{T}\phi^-(x)\phi^-(x')}
    \end{bmatrix},\label{TheGreenFunction}
\eeq
where $T$ ($\Bar{T}$) denote (anti-)time ordering, satisfying
\begin{align}
    \av{T\phi(x)\phi(x')} &= \Theta(t-t')\av{\phi(x)\phi(x')}+\Theta(t'-t)\av{\phi(x')\phi(x)}, \\
    \av{\Bar{T}\phi(x)\phi(x')} &= \Theta(t-t')\av{\phi(x')\phi(x)}+\Theta(t'-t)\av{\phi(x)\phi(x')}.    
\end{align}
The two-point functions are related via 
\beq
     G^{++}(x,x') +G^{--}(x,x') = G^{+-}(x,x') +G^{-+}(x,x').
\eeq
For classical considerations it is useful to work in the Keldysh basis, in which the Schwinger basis fields $\phi^\pm$ are transformed into ''classical'' and ''quantum'' fields 
$\phi^\text{c}$,  $\phi^\text{q}$ via a transformation matrix $U$ as
\beq
    \Phi(x)\rightarrow U\Phi(x)= 
    \begin{bmatrix}
        \tfrac{1}{2}[\phi^+(x)+\phi^-(x)]\\\phi^+(x)-\phi^-(x)
    \end{bmatrix}
    \equiv 
    \begin{bmatrix}
        \phi^\text{c}(x)\\\phi^\text{q}(x)
    \end{bmatrix},\qquad U=\begin{bmatrix}
        \tfrac{1}{2} & \tfrac{1}{2} \\ 1 & -1
    \end{bmatrix}.
\eeq
This notation can be understood heuristically in the sense that $\phi^\text{q}$ expresses the amplitude of the quantum fluctuations around the mean field value $\phi^\text{c}$.
The kinetic operator in the Keldysh basis is
\beq
   \mathbbm{G}^{-1}(x,x') \rightarrow U\mathbbm{G}^{-1}(x,x')U^T = \begin{bmatrix}
    0& G^{-1}(x)  \\ G^{-1}(x)& 0
    \end{bmatrix}\delta(x,x'), 
    \label{eq:keldyshinvG}
\eeq
and the propagators are
\beq
     \mathbbm{G}(x,x')  = 
     \begin{bmatrix}
    -iG_F(x,x') & G_R(x,x') \\ G_A(x,x') & 0
    \end{bmatrix},
    \label{eq:KeldyshProp}
\eeq
with 
\beq
    -iG_F(x,x') &= \tfrac{1}{2}\left[G^{+-}(x,x')+G^{-+}(x,x')\right] = -\tfrac{i}{2}\av{\{\phi(x),\phi(x')\}},\\
    G_R(x,x') &= G^{++}(x,x')-G^{+-}(x,x') =-i\Theta(t-t')\av{[\phi(x),\phi(x')]},\\
    G_A(x,x') &= G^{++}(x,x')-G^{-+}(x,x') =-i\Theta(t'-t)\av{[\phi(x'),\phi(x)]}.
    \label{eq:propConstituents}
\eeq
We recognise the statistical Feynman propagator $G_F$ and spectral retarded/advanced propagators $G_R(x,x')=G_A(x',x)$. 
The free propagators satisfy the equations of motion
\beq
    G^{-1}(x)G_F(x,x')=0,\qquad
    G^{-1}(x)G_{R,A}(x,x')=\frac{\delta(x-x')}{\sqrt{-g(x)}}.
    \label{eq:freePropEoMs}
\eeq

\subsection{Coarse graining and window functions}
\label{sec:coarsegraining}

In order to derive the effective IR dynamics of the inflaton field $\phi$, one may proceed to split it into UV and IR parts as
$\phi=\phi_\uv+\phi_\ir$. 
The field split has been addressed at the level of the effective action in earlier work \cite{Morikawa1990} 
(see also \cite{Matarrese:2003ye,PerreaultLevasseur:2013kfq}), 
where an effective IR equation of motion is obtained by integrating out the UV modes.

The IR/UV field split is a coarse-graining procedure, in the sense that it amounts to averaging in position space using a smoothing window function $\overline{W}$ dependent on a characteristic smoothing scale $L$ as \cite{Winitzki:1999ve}
\beq
    \phi_\ir(x)=\int\dd^3\mathbf{x'}\phi(x')\overline{W}(\mathbf{x}-\mathbf{x'},L).
    \label{eq:IRfieldDef}
\eeq
The IR field value at a given point in space is then given by the value of the full field $\phi$ averaged over a region of size $L$.
The convolution of the window function in position space then corresponds to multiplication in momentum space. Implicit in \eqref{eq:IRfieldDef} is that $\overline{W}$ is taken to be local in time, 
through a form
\beq
    \overline{W}(x,x') = \delta(t-t')\int\frac{\dd^3\mathbf{k}}{(2\pi)^3}\overline{W}(k,t)e^{i\mathbf{k}\cdot(\mathbf{x}-\mathbf{x'})}.
\eeq
Note that $k$ is the co-moving momentum, with $k/a(t)$ being the physical momentum. We will use this form in the following as well. From the field split $\phi=\phi_\uv+\phi_\ir$ with $\phi_\ir$ defined as in \eqref{eq:IRfieldDef}, it follows that we can define the UV part of the field in terms of a completing function $W$, 
\beq
    W(x,x')+\overline{W}(x,x')=\delta(x-x'),
\eeq
such that 
\beq
    \phi_\uv(x)=\int_{x'}W(x,x')\phi(x'), \qquad \phi_\ir(x)=\int_{x'}\overline{W}(x,x')\phi(x').
    \label{eq:UVfieldSplit}
\eeq
In order for the window function to truly split the field into an IR and UV part, we require them to satisfy $W(k,t)=1$ for field modes with $k\gg aH$ and $W(k,t)=0$ for $k\ll aH$, and vice versa for $\overline{W}(k,t)$.
In particular, if $W(k,t)$ has precisely the value 1 or 0 for a given mode, that entire mode belongs to either the UV or IR field. For a smooth window function interpolating between 1 and 0, modes of all $k$ will contribute to both the IR and the UV field. In that case, the field split is into a "mostly-UV" and "mostly-IR" part.

We will perform the split in terms of physical momentum $k/a(t)$, rather than the co-moving momentum defining the mode functions. In practice, this introduces a time-dependence of the window function, as the UV modes are continuously red-shifted and  join the coarse-grained classical IR modes. For the IR modes the effect of these quantum fluctuations crossing the Hubble horizon is what amounts to a continuous noise source. 

\subsection{The UV vacuum mode functions}
\label{sec:modefunctions}

Expanding the propagator in terms of quantised field momentum modes 
\beq
    \phi(x) = \int\frac{\dd^3\mathbf{k}}{(2\pi)^3}\big[ a_\mathbf{k}\phi(k,t)e^{i\mathbf{k}\cdot\mathbf{x}}+\text{h.c.}\big],
\eeq
with $[a_\mathbf{k},a_\mathbf{k'}^\dagger]=(2\pi)^3\delta^3(\mathbf{k}+\mathbf{k'})$ and $a_\mathbf{k}$ annihilating the Bunch-Davies vacuum, the free equation of motion \eqref{eq:freePropEoMs} has the standard solution
\beq
  \phi(k,t)=\frac{\sqrt{\pi}}{2\sqrt{a^3H(1-\epsilon_H)}}H_\nu^{(1)}\left(\frac{k}{aH(1-\epsilon_H)}\right),\, 
    \label{eq:SRdSmodesolution}
\eeq
where 
$\nu^2=\frac{9}{4}-3\epsilon_M+3\epsilon_H$ and $H_\nu^{(1)}(x)$ are the Hankel functions of the first kind \cite{grad}. In the massless dS case, $\nu=\frac{3}{2}$ and the mode solution reduces to
\beq
    \phi(k,t)=-\frac{H}{\sqrt{2k^3}}\bigg(i+\frac{k}{aH}\bigg)e^{ik/aH}, \qquad \epsilon_M=0,\quad \epsilon_H=0.
    \label{eq:dSmodesolution}
\eeq
In the long-wavelength limit, the Hankel function can be approximated\footnote{For small $x$ and $\nu>0$ the leading terms are
$H_{\nu}^{(1)}(x)\simeq -i\frac{\Gamma[\nu]}{\pi}\left(\frac{x}{2}\right)^{-\nu}\left[1+\frac{1}{\nu-1}\left(\frac{x}{2}\right)^2+\dotso\right].$}
such that the leading term in $k/aH$ of the mode function \eqref{eq:SRdSmodesolution} is given by
\beq
    \phi(k,t) \simeq 
    -i\frac{H(1-\epsilon_H)}{\sqrt{2k^3}}\left(\frac{k}{aH(1-\epsilon_H)}\right)^{\epsilon_M-\epsilon_H}, \qquad \epsilon_M\ll 1, \quad k/aH\ll 1.
    \label{eq:IRappmode}
\eeq
In the following, whenever the quantum UV modes enter through their correlators, these will be represented by the free vacuum state through the above mode functions (\ref{eq:SRdSmodesolution})--(\ref{eq:IRappmode}). The long-wavelength approximate solution \eqref{eq:IRappmode} applies to our UV modes, since as we will see below, the modes responsible for the stochastic noise are in fact super-horizon, in the sense required by (\ref{eq:IRappmode}). They are modes transitioning from the UV to the IR.

\paragraph{Interacting UV mode functions.}

In order to solve explicitly for the UV mode functions, it is usually assumed that interactions may be neglected. It is however worth noting, that since the mass only enters through $\epsilon_M$, the expressions generalise trivially to Gaussian interacting UV modes including an effective mass $M^2=m^2 + \dotso$. In the complete theory, where also the UV modes experience an effective mass $\propto \sqrt{\lambda}H^2$, modes are never truly massless, and for $m^2=0$, the small-mass criterion $\epsilon_M\ll 1$ becomes a constraint on the coupling, $\sqrt{\lambda}\ll 1$.

\subsection{IR effective theory}
\label{sec:IReffective}
In the language of open quantum systems, the quantum UV modes can be viewed as a bath affecting the system of classical IR modes. 
The generating functional \eqref{eq:genfungeneral} can then be written in terms of an influence functional $\mathcal{F}$ as
\beq
    Z[\mathbbm{J}] = \int\mathcal{D}\Phi_\ir e^{i(S[\Phi_\ir]-\mathbbm{J}^T\Phi_\ir)}\mathcal{F}[\Phi_\ir,\mathbb{J}],
    \label{eq:fullaction}
\eeq
where 
\beq
    \mathcal{F}[\Phi_\ir,\mathbb{J}]=\int \mathcal{D}\Phi_\uv\exp{i\int_x\left(\tfrac{1}{2}\Phi_\uv^T\mathbbm{G}^{-1}\Phi_\uv
    +\Phi_\ir^T\mathbbm{G}^{-1}\Phi_\uv-V_\text{I}(\Phi_\ir,\Phi_\uv)+\mathbbm{J}^T\Phi_\uv\right)},
    \label{eq_InfluenceFunctional}
\eeq
contains all the instances of $\Phi_\uv$ and is to become an effective contribution the IR dynamics upon integrating out the UV part of the field. Here we have defined $V_\text{I}(\Phi_\ir,\Phi_\uv)\equiv V_\text{I}(\Phi_\ir+\Phi_\uv)-V_\text{I}(\Phi_\ir)$, which typically includes non-linear self-interactions among the UV modes, and between the UV and IR modes. 
We are interested in the stochastic noise that arises solely from the time-dependent field split of the free theory, and we will therefore simply neglect this non-linear interaction in our calculation, and as in the preceding section consider only free UV fields.\footnote{Stochastic contributions from the non-linear IR-UV interactions involve powers of a coupling which for inflation tend to be small.} The self-interaction of the IR field is retained in (\ref{eq:fullaction})
Setting also
$\mathbb{J}=0$ in \eqref{eq_InfluenceFunctional}, we may complete the square and perform the path integral over $\Phi_\uv$, to obtain
\beq 
    \mathcal{F}_0[\Phi_\ir] = N\exp{-\frac{i}{2}\int_x\int_{x'}\Phi_\ir^T(x)\mathbbm{G}^{-1}(x)\mathbbm{G}_\uv(x,x')\mathbbm{G}^{-1}(x')\Phi_\ir(x')}, \label{eq_UVinfluenceFunc}
\eeq
where $N\propto(\det \mathbbm{G}_\uv^{-1})^{-1/2}$ is a normalisation factor and UV propagator is defined by \eqref{eq:KeldyshProp} with \eqref{eq:UVfieldSplit} as
\beq
    \mathbbm{G}_\uv(x,x')=W(x)\mathbbm{G}(x,x')W(x'),
    \label{eq:fieldsplitUVprop}
\eeq
Here $G^{-1}(x')$ can be integrated by parts to act on ${G}_\uv(x,x')$ rather than the field $\Phi_\ir(x')$. 
Later we will also consider departures from strict dS space ($\epsilon_H\neq 0$), but for the moment we continue with simply
\beq
    G^{-1}(x)=-\partial_t^2-3H\partial_t+\frac{\nabla^2_\mathbf{x}}{a^2(t)}-m^2.
\eeq
Writing out the kernel in \eqref{eq_UVinfluenceFunc} in the Keldysh c-q matrix form, we explicitly have
\beq
\mathcal{F}_0[\Phi_\ir] &= N\mathrm{exp}\bigg[ -\frac{i}{2}\int_x\int_{x'}\bigg(
    -i[\phi_\ir^\text{c}(x),\phi_\ir^\text{q}(x)]
    \int\frac{\dd^3\mathbf{k}}{(2\pi)^3}e^{i\mathbf{k}\cdot(\mathbf{x}-\mathbf{x'})}\\
    &\quad\times\left[\begin{smallmatrix}
        0 & [{\mathcal{Q}}_t+W_tG^{-1}_t][{\mathcal{Q}}_{t'}+W_{t'}G^{-1}_{t'}]G_A(t,t')\\
        [{\mathcal{Q}}_t+W_tG^{-1}_t][{\mathcal{Q}}_{t'}+W_{t'}G^{-1}_{t'}]G_R(t,t') & -i{\mathcal{Q}}_t{\mathcal{Q}}_{t'}G_F(t,t')
    \end{smallmatrix}\right]
    \bigg[\,\begin{matrix}\phi_\ir^\text{c}(x')\\\phi_\ir^\text{q}(x')\end{matrix}\,\bigg]
    \bigg)\bigg],
    \label{eq:infFunc1}
\eeq
where the UV propagator components $G_F$, $G_{R/A}$, $G^{-1}$ and the window function $W$ enter through their momentum space counterparts, and we have defined the operator
\beq
{\mathcal{Q}}_x = \int\frac{\dd^3{\bf k}}{(2\pi)^3} {\mathcal{Q}}_t(k)e^{i\mathbf{k}\cdot(\mathbf{x}-\mathbf{x'})},\quad   {\mathcal{Q}}_t(k)\equiv-\Ddot{W}(k,t)-3H\dot{W}(k,t)-2\dot{W}(k,t)\partial_t.
\eeq
As we will argue below, the off-diagonal c-q terms in \eqref{eq:infFunc1} may under some circumstances be neglected compared to the diagonal q-q component. 
Since the UV fields are taken to be non-interacting, we can make use of \eqref{eq:freePropEoMs}, and after partially integrating with respect to time, one may write for the off-diagonal components
\beq
    &\int_x\int_{x'}\phi_\ir^\text{q}(x)\phi_\ir^\text{c}(x')[{\mathcal{Q}}_x+W_xG^{-1}_x][{\mathcal{Q}}_{x'}+W_{x'}G^{-1}_{x'}]G_R(x,x')\\
    &=\int_x\int_{x'}\phi_\ir^\text{q}(x)\left[\Upsilon(x,x')
    +\frac{\delta(t-t')}{a^3(t)}\left(W_{x'}G^{-1}_{x'}
    +2\mathcal{Q}_{x'}W_{x'}+2W_{x'}\dot{W}_{x'}\partial_{t'}\right)\right]\phi_\ir^\text{c}(x'),
    \label{eq:offdiag}
\eeq
where we have defined the quantity
\beq
    \Upsilon(x,x')\equiv\int\frac{\dd^3\mathbf{k}}{(2\pi)^3}{\mathcal{Q}}_t{\mathcal{Q}}_{t'}G_R(k,t,t')e^{i\mathbf{k}\cdot(\mathbf{x}-\mathbf{x'})}.
    \label{eq:UpsilonDef}
\eeq
The last three terms in \eqref{eq:offdiag} include a window function $W$ acting on an IR-field. As pointed out in \cite{Morikawa1990}, if $W(k,t)$ is a projection operator (so that $W^2\phi=W\phi$, $\overline{W}W\phi=0$), terms for which the IR field $\phi_\ir=\overline{W}\phi$ is directly convoluted with $W$ vanish. 

Requiring window functions to be projections is a rather strict constraint, since it requires $W$ to take on only the values 0 and 1, and hence be a (sequence of) discontinuous step functions. 
A smooth window function is not a projection, but one may still expect that terms of the form $W\phi_{\rm IR}$ are suppressed also for e.g. a smoothed-out step function. Discarding the last three terms of \eqref{eq:offdiag} then implies that the window function is assumed to be "sufficiently step-like". In the following we will neglect these terms, but keep in mind this requirement on the window function. The final off-diagonal term $\Upsilon$ does not contain any $W$ that gets directly convoluted with the $\phi_\ir^\text{c}, \phi_\ir^\text{q}$ fields. In section \ref{sec:checkingGR} we will show by an explicit computation that $\Upsilon$ is subleading compared to the q-q component in \eqref{eq:infFunc1}, but until then it is kept in the remainder of this section.

When the simplifications discussed above can be made, the influence functional can be written as
\beq
    \mathcal{F}_0[\Phi_\ir] = N\mathrm{exp}\bigg[ i\int_x\int_{x'}\bigg(\frac{i}{2}\phi_\ir^\text{q}(x)\Re\Pi(x,x')\phi_\ir^\text{q}(x')-2\Theta(t-t')\phi_\ir^\text{q}(x)\Im\Pi(x,x')\phi_\ir^\text{c}(x') \bigg)\bigg],
    \label{eq:influenceFuncSimplifed}
\eeq
with 
\beq
    \Pi(x,x') = \int\frac{\dd^3\mathbf{k}}{(2\pi)^3}
    {\mathcal{Q}}_t{\mathcal{Q}}_{t'}\phi(k,t)\phi^*(k,t')e^{i\mathbf{k}\cdot(\mathbf{x}-\mathbf{x'})},
    \label{eq:Pixx}
\eeq
having used that (for $t>t'$)
\beq
    G_F(k,t,t')=\Re \phi(k,t)\phi^*(k,t'), \qquad G_R(k,t,t')=2\Theta(t-t')\Im \phi(k,t)\phi^*(k,t').
\eeq
In order to compute these quantities explicitly, one is required to choose a concrete representation for the UV field mode functions $\phi(k,t)$ as well as a window function $W(k,t)$. To this end we will be using the mode solutions defined in section \ref{sec:modefunctions}, at different levels of approximation.

\subsection{Stochastic IR theory}
\label{sec:IRstochastic}
Following a well-known procedure for turning our effective theory into a stochastic one \cite{Stratonovich,Hubbard}, the q-q component of \eqref{eq:influenceFuncSimplifed} is represented by a real-valued quantity by introducing an auxiliary field $\xi(x)$ via a Hubbard-Stratonovich transformation,
\beq
    e^{-\frac{1}{2}\int_x\mathcal{M}\phi_\ir^2} = \int\mathcal{D}\xi e^{-\frac{1}{2}\int_x\mathcal{M}^{-1}\xi^2+i\int_x\xi \phi_\ir},
\eeq
for which we obtain
\beq 
    \mathcal{F}_0[\Phi_\ir] = N\int\mathcal{D}\xi\,
    \mathrm{exp}\bigg[\int_x\int_{x'}&\Big(-\tfrac{1}{2}\xi(x)\Re \Pi(x,x')^{-1}\xi(x')+ i\xi(x)\phi_\ir^\text{q}(x')\\
    &\qquad-i2\Theta(t-t')\phi_\ir^\text{q}(x)\Im\Pi(x,x')\phi_\ir^\text{c}(x')\Big)\bigg].
    \label{InflFuncRes}
\eeq
Here the variables $\xi$ are defined to be Gaussian, with 
\begin{align}
    \av{\xi(x)\xi(x')}= \int\mathcal{D}\xi\mathcal{P}[\xi]\xi(x)\xi(x')=\Re \Pi(x,x').
    \label{eq:noisecorr}
\end{align}
In this relation $\av{\cdot}$ denotes the ensemble average and the right-hand-side is evaluated as a quantum expectation value.

An equivalent, but perhaps more familiar form of the stochastic noise correlator \eqref{eq:noisecorr} can be obtained by rewriting  \eqref{eq:Pixx} into\footnote{Using the metric determinant to write $-a^3(t){\mathcal{Q}}_t\phi(k,t) = \partial_t[a^3(t)\dot{W}(k,t)\phi(k,t)]+a^3(t)\dot{W}(k,t)\dot{\phi}(k,t)$ and partial integrating the first term in the action.}
\beq
    \Re\int_x\int_{x'}\phi_\ir^\text{q}(x)\Pi(x,x')\phi_\ir^\text{q}(x') = \Re\int_x\int_{x'} \int\frac{\dd^3\mathbf{k}}{(2\pi)^3}
    \left(-\dot{\phi}^\text{q}_{\ir}(t)\phi(t)+\phi^\text{q}_{\ir}(t)\dot{\phi}(t)\right) \dot{W}_t&\\
    \times\dot{W}_{t'}\Big(-\dot{\phi}^\text{q}_{\ir}(t') \phi^*(t')+\phi^\text{q}_{\ir}(t')\dot{\phi}^*(t')\Big)e^{i\mathbf{k}\cdot(\mathbf{x}-\mathbf{x'})}&, 
\eeq
and introducing instead two auxiliary fields $\xi_\phi$ and $\xi_\pi$. Writing $\pi=\dot{\phi}$, the generating functional becomes 
\beq
\mathcal{F}_0[\Phi_\ir] &= N\int\mathcal{D}\xi_\phi\mathcal{D}\xi_\pi\,
    \mathrm{exp}\bigg[\int_x\int_{x'}\bigg(-\tfrac{1}{2} \left[\,\xi_\phi(x),\xi_\pi(x)\,\right]\mathcal{M}^{-1}(x,x')\bigg[\,\begin{matrix}\xi_\phi(x')\\\xi_\pi(x')\end{matrix}\,\bigg]\\
    &\quad+ i\left[\,-\pi_\ir^\text{q}(x),{\phi}_\ir^\text{q}(x)\,\right]\bigg[\,\begin{matrix}\xi_\phi(x')\\\xi_\pi(x')\end{matrix}\,\bigg]
    -i2\Theta(t-t')\phi_\ir^\text{q}(x)\Im\Pi(x,x')\phi_\ir^\text{c}(x')\bigg)\bigg],
    \label{eq:influencefinal}
\eeq
where
\beq
    \mathcal{M}(x,x')
    = \Re \int\frac{\dd^3\mathbf{k}}{(2\pi)^3}e^{i\mathbf{k}\cdot(\mathbf{x}-\mathbf{x'})}
    \dot{W}_t
    \begin{bmatrix}
        \phi(k,t)\phi^*(k,t') & \phi(k,t)\pi^*(k,t')\\
        \pi(k,t)\phi^*(k,t') & \pi(k,t)\pi^*(k,t')
    \end{bmatrix}
    \dot{W}_{t'}.
    \label{eq:noisematrix}
\eeq
Schematically, we may then write for the stochastic noise 
\beq
    \xi_\phi(x) &= \int\frac{\dd^3\mathbf{k}}{(2\pi)^{\frac{3}{2}}}\dot{W}_t\big[a_{\bf k}\phi(k,t)e^{i\mathbf{k}\cdot\mathbf{x}}+\textrm{h.c.}\big],\\
    \xi_\pi(x) &= \int\frac{\dd^3\mathbf{k}}{(2\pi)^{\frac{3}{2}}}\dot{W}_t \big[a_{\bf k}\pi(k,t)e^{i\mathbf{k}\cdot\mathbf{x}}+\textrm{h.c.}\big], \label{schemNoiseExprs}
\eeq
keeping in mind that $\xi(x)$ is a number and the right-hand side is an operator, and the correspondence is at the level of expectation values. 

\paragraph{The stochastic equation.}

Combining the expressions \eqref{eq_InfluenceFunctional} and \eqref{InflFuncRes}, the influence functional \eqref{eq:fullaction} now reads 
\beq
    Z[0]=\int \mathcal{D}\Phi_\ir e^{iS[\Phi_\ir]}\mathcal{F}_0[\Phi_\ir]=\int \mathcal{D}\Phi_\ir\mathcal{D}\xi e^{iS_{\rm eff}}[\Phi_\ir,\xi],
\eeq
with
\beq
    S[\Phi_\ir] = \int_x \left(\tfrac{1}{2}\phi_\ir^\text{c} G^{-1}\phi_\ir^\text{q} - V_\text{I}[\phi_\ir^\text{c},\phi_\ir^\text{q}]\right).
\eeq
By variation of the effective action, we finally arrive at the stochastic equation of motion for the IR field, which becomes\footnote{We note that the potential term $V_\text{I}[\Phi]=V_\text{I}[\phi^+]-V_\text{I}[\phi^-]$ in the Keldysh basis has the property that $\dd V_\text{I}[\Phi]/\dd \phi^\text{q}|_{\phi^\text{q}=0}= V_\text{I}'[\phi^\text{c}]$.}
\beq
    0=\frac{\delta S_{\rm eff}}{\delta \phi_\ir^\text{q}(x)}\bigg|_{\phi_\ir^\text{q}=0}
    &=\ddot{\phi}_\ir^\text{c}(x)+3H\dot{\phi}_\ir^\text{c}(x)-\frac{\nabla^2_\mathbf{x}}{a^2(t)}\phi_\ir^\text{c}(x)+V'(\phi_\ir^\text{c}(x))- \xi(x)\\
    &\qquad+\int_{x'}2\Theta(t-t')\Im\Pi(x,x')\phi_\ir^\text{c}(x').
    \label{eq:LangevinEoM}
\eeq
One may note that if the IR field is sufficiently super-horizon, the gradient term is negligible, since by construction
\beq
\frac{k^2}{a^2(t)H^2}\phi_\ir^\text{c}\ll \phi_\ir^\text{c}.
\eeq 
This must then be compared to the remaining terms.
To the extent that $\Upsilon(x,x') = 2\Theta(t-t')\Im\Pi(x,x')$ may be neglected, we recover the stochastic equation advertised in (\ref{eq:twonoise});
\beq
    \ddot{\phi}_\ir+3H\dot{\phi}_\ir+V'(\phi_\ir)=\xi.
\eeq
One may equivalently vary instead \eqref{eq:influencefinal}, to obtain the identification
\beq
     3H\xi_\phi(x)+\dot{\xi}_\phi(x)+\xi_\pi(x)\equiv\xi(x),
     \label{eq:decomposedNoises}
\eeq
as discussed in section \ref{sec:langevin}. In the following sections we will show explicitly that in certain limits, $3H\xi_\phi$ is indeed the dominant contribution to the noise. For the moment, we will simply note that for $k\ll aH$ the mode derivative is $\dot{\phi}(k,t)\simeq \epsilon_M H \phi(k,t)$, as can be seen from e.g. \eqref{schemNoiseExprs} and \eqref{eq:BigQ}. It follows that for small $\epsilon_M$, the remaining correlators in \eqref{eq:noisematrix} are $\mathcal{O}(\epsilon_M)$.

Under the above assumptions, and additionally in the slow-roll regime where $\ddot{\phi}\ll3H\dot{\phi}$, the original stochastic inflation form (\ref{eq:Langevin1}) is recovered, 
\beq
    \dot{\phi}_\ir(t)+\frac{V'(\phi_\ir(t))}{3H} = \xi_\phi(t).
\eeq
Let us summarise the procedure so far. Before applying standard slow-roll and large wavelength assumptions common to inflationary dynamics, we have ignored UV-UV and UV-IR self-interactions, made use of $W$ being close to a projection onto UV modes, neglected $\Upsilon$ and neglected noise contributions other than $\xi_\phi$. The IR fields have also implicitly been assumed classical, so that the variation of the path integral with respect to $\phi_\ir^\text{q}$ gives the complete dynamics. Only then do we recover the standard form of the stochastic equation. A number of these assumptions will be checked explicitly below. 

Before we proceed, we emphasise that expressing the path integral in terms of a stochastic process does not imply any physical reality to each individual stochastic trajectory. Only the ensemble average over initial conditions and realisations of the noise have meaning, and so conceptually, the corresponding Fokker-Planck equation and distribution is perhaps the preferred object to work with.

\subsection{An example: the step window function}
\label{sec:Starobinsky}

The simplest example of a window function is the step function
\beq
    W(k,t)={\Theta(k/aH-\mu)},
    \label{eq:stepfunction}
\eeq
which projects out modes with $k<\mu aH$, and was the original choice made in \cite{Starobinsky:1986fx,Starobinsky:1994bd}. It is a true projection, and so some of the simplifications discussed above go through straightforwardly. The dimensionless parameter $\mu<1$ is taken to be small but non-vanishing, and the coarse-graining scale is then $(\mu aH)^{-1}$.

The noise correlations can be computed from \eqref{eq:noisematrix}, where from \eqref{eq:IRappmode} we obtain to first order in mass and slow-roll parameters, 
\beq
    \begin{bmatrix}
    \av{\xi_\phi(x)\xi_\phi(x')}\\
    \av{\xi_\phi(x)\xi_\pi(x')}\\
    \av{\xi_\pi(x)\xi_\pi(x')}
    \end{bmatrix}
    &= \frac{H^3}{4\pi^2}(1-\epsilon_H)^3\frac{\sin \mu aHr}{\mu aHr}\delta(t-t')\bigg(\frac{\mu}{1-\epsilon_H}\bigg)^{2\epsilon_M-2\epsilon_H} \\
    &\qquad\times
    \begin{bmatrix}
    1+\mu^2(1+2\epsilon_M)\\
    \left(-\epsilon_M+\mu^2(1+\tfrac{3}{2}\epsilon_M-\epsilon_H)\right)H\\
    \left(\epsilon_M^2+2\mu^2\epsilon_M\right)H^2
    \end{bmatrix}+\mathcal{O}(\mu^4).
    \label{eq:StarobNoise2}
\eeq
In the limit $\epsilon_M,\epsilon_H\rightarrow0$ the first noise correlator in \eqref{eq:StarobNoise2} reduces to the original result \cite{Starobinsky:1994bd} to leading order in $\mu$. It remains to take the limits $\mu\ll 1$ and $r<aH$ to recover the correlation,
\beq
    \av{\xi_\phi(x)\xi_\phi(x')} = \frac{H^3}{4\pi^2}\delta(t-t').
    \label{eq:StarobNoise3}
\eeq
This result is elegant in its simplicity, but as we have seen, it relies on a number of approximations and assumptions. An early discussion of noise resulting from non-linear interactions between the IR-UV fields with a  cutoff scale separation includes \cite{Hu:1992ig}.

The expression \eqref{eq:StarobNoise2} is our first encounter 
of the correction function $f(\epsilon_M,\epsilon_H,\mu, ...)$ advertised in section \ref{sec:langevin}. We see that the limits $\epsilon_M,\epsilon_H\rightarrow 0$ do not commute with $\mu\rightarrow 0$. Indeed, as soon at $\mu$ is taken to be small, the standard noise normalisation (\ref{eq:StarobNoise3}) only follows in the strict massless and dS limit. 
From \eqref{eq:StarobNoise2} it is also apparent that the contribution of $\xi_\phi$ dominates for small $\mu$.

It was pointed out in \cite{Winitzki:1999ve} that some window functions fail to reproduce the correct long-distance behaviour of the full (non-coarse grained) correlators, and that some result in coloured noise \cite{Casini:1998wr,Winitzki:1999ve}.\footnote{For instance, the resulting correlations $\av{\dot{\phi}_\ir(x)\dot{\phi}_\ir(x')}$ from the Langevin equation should at least match the behaviour of  $\av{\dot{\phi}(x)\dot{\phi}(x')}\sim r^{-4}$ at large distances. This can be achieved by demanding sufficient smoothness of the window function, such that the noise correlation 
\beq
    \av{\xi_\phi(x)\xi_\phi(x')}\sim \int\dd kh(k)\sin kr = \frac{h(0)}{r}-\frac{h''(0)}{r^3}+\frac{h'''(0)}{r^5}-\dotso,
\eeq
has at least $h^{\prime\prime\prime}(0)$ finite. The authors of \cite{Winitzki:1999ve} provided a sufficient (but not necessary or exhaustive) condition for a window function to satisfy this.} The step function is one such example, and in the following, we will consider smooth window functions.

\subsection{Momentum decomposition of the propagator}
\label{sec:propagatordecomp}

An alternative method for the scale separation between IR and UV modes was applied in \cite{Moss:2016uix}, where the propagator rather than the field itself is subject to a window function in momentum space. 
The propagator is decomposed into two constituents that are dominant in the IR and UV respectively:
\beq 
    \mathbbm{G}(x,x')=\mathbbm{G}_\ir(x,x')+\mathbbm{G}_\uv(x,x'),
    \label{eq:propsplit}
\eeq 
where the IR propagator is weighted according to
\beq 
    \mathbbm{G}_\ir(x,x') \equiv \int_{y}\int_{y'}\overline{W}(x,y)\mathbbm{G}(y,y')\overline{W}(y',x'), \label{UVProp}
\eeq
and $\overline{W}$ is as earlier used to project the field onto long wavelengths.
The UV propagator is constructed from \eqref{eq:propsplit} as (c.f. (\ref{eq:fieldsplitUVprop}))
\beq 
    \mathbbm{G}_\uv(x,x') = \int_{y'}\int_y\left[\mathbbm{G}(x,y)W(y,x')+W(x,y)\mathbbm{G}(y,x')-W(x,y)\mathbbm{G}(y,y')W(y',x')\right].
    \label{eq:GUVrigDef}
\eeq
An additional step is to make the formal rewriting of the field into two constituents,
\beq 
    \Phi(x) = \Phi_1(x)+\Phi_2(x).
    \label{eq:phi1phi2}
\eeq
Unlike when decomposing at the level of the fields, these field constituents are not under any constraints: they are simply two new field variables defined over all of momentum space. The decomposition \eqref{eq:phi1phi2} is used as a tool together with the propagator constituents \eqref{eq:propsplit} to make use of a Gaussian identity\footnote{
    For some function $f(x)$ depending only on the combination $x=y+z$ the relation
    \beq
        \sqrt{\frac{|a+b|}{2\pi ab}}\int\dd y\int\dd z\exp{-\left(\frac{y^2}{2a}+\frac{z^2}{2b}+f(y+z)\right)}=\int\dd x \exp{-\left(\frac{x^2}{2(a+b)}+f(x)\right)},
    \eeq
    can be verified e.g. by a change of variables $x=y+z$ and integration over $u=by-az$. }
in order to rewrite the generating functional \eqref{eq:genfungeneral} as 
\beq 
    Z[\mathbbm{J}] = \int\mathcal{D}\Phi_1\mathcal{D}\Phi_2\exp{\frac{i}{2}\int_x \left(\Phi_1^T\mathbbm{G}^{-1}_\ir\Phi_1+\Phi_2^T\mathbbm{G}^{-1}_\uv\Phi_2-2V_\text{I}(\Phi_1+\Phi_2)+\mathbbm{J}^T(\Phi_1+\Phi_2)\right)}. \label{eq:GenFuncRigForm}
\eeq
We have in mind a smooth window function $W$($\overline{W}$) that suppresses the IR(UV). From the construction of the propagators, the UV components of the $\Phi_1$ field will have suppressed contributions to the path integral, and analogously the IR components of the $\Phi_2$ field. Note that when decomposing at the level of the propagator, we no longer have bilinear kinetic terms mixing the UV and IR physics in the generating functional. 
To obtain an expression for the IR kinetic operator in \eqref{eq:GenFuncRigForm}, we write (formally, suppressing integration labels)
\beq
    \mathbbm{G}_\ir^{-1} = \frac{1}{1-\mathbbm{G}^{-1}\mathbbm{G}_\uv}\mathbbm{G}^{-1}.
    \label{eq:Gexpand}
\eeq
For the large-wavelength dynamics governed by $\mathbbm{G}_\ir^{-1}$, the propagator $\mathbbm{G}_\uv$ is by construction acting on a subspace of the field configuration in which it is suppressed, hence allowing the expansion
\beq
    \mathbbm{G}_\ir^{-1} &= \mathbbm{G}^{-1}+\mathbbm{G}^{-1}\mathbbm{G}_\uv\mathbbm{G}^{-1}+\mathbbm{G}^{-1}\mathbbm{G}_\uv\mathbbm{G}^{-1}\mathbbm{G}_\uv\mathbbm{G}^{-1}+\dotso . 
    \label{eq:InversePropExpansion}
\eeq
To clarify, the "small quantity" in the expansion is the degree to which the "mostly-UV" propagator $\mathbbm{G}_\uv$ is small when acting on the "mostly-IR" field $\Phi_1$.\footnote{For a strict step function such as (\ref{eq:stepfunction}), this procedure is ill-defined, since the denominator of (\ref{eq:Gexpand}) vanishes for UV modes.} For a given window function, it may be prudent to confirm convergence a posteriori, and we will do this below.

The first term in \eqref{eq:InversePropExpansion} replaces $G_\ir^{-1}$ by $G^{-1}$ in the quadratic part of the action \eqref{eq:GenFuncRigForm}, such that
\beq
    Z[0] &= \int \mathcal{D}\Phi_1 
    \exp{\frac{i}{2}\int_x \left(\Phi_1^T\mathbbm{G}^{-1}\Phi_1-2V_\text{I}(\Phi_1)\right)}\\
    &\quad\times \exp{\frac{i}{2}\int_x\Phi_1^T\mathbbm{G}^{-1}\mathbbm{G}_\uv\mathbbm{G}^{-1}\Phi_1+\Phi_1^T\mathbbm{G}^{-1}\mathbbm{G}_\uv\mathbbm{G}^{-1}\mathbbm{G}_\uv\mathbbm{G}^{-1}\Phi_1+\dotso}\\
    &\quad\times\int\mathcal{D}\Phi_2\exp{\frac{i}{2}\int_x \left(\Phi_2^T\mathbbm{G}^{-1}_\uv\Phi_2-2V_\text{I}(\Phi_1,\Phi_2)\right)}.
\eeq
where similarly to \eqref{eq_InfluenceFunctional}, we have defined $V_\text{I}(\Phi_1,\Phi_2)\equiv V_\text{I}(\Phi_1+\Phi_2)-V_\text{I}(\Phi_1)$. 

By the same logic as when the UV-IR and UV-UV self-interactions were neglected in the previous section, we will treat the $\Phi_2$ field as non-interacting, in which case the integral over $\Phi_2$ is Gaussian and only provides a multiplicative constant. The second line will now play the role of the influence functional, while the first line is the free action $S[\Phi_1]$.

Our new influence functional has as its leading term (corresponding to the next-to-leading order (NLO) in the inverse propagator expansion \eqref{eq:InversePropExpansion})
\beq
    \mathbbm{G}^{-1}\mathbbm{G}_\uv\mathbbm{G}^{-1} = W\mathbbm{G}^{-1}+\mathbbm{G}^{-1}W-\mathbbm{G}^{-1}W\mathbbm{G}W\mathbbm{G}^{-1}, \label{NLOInverseProp}
\eeq
which includes the integrand in \eqref{eq_UVinfluenceFunc}. The derivation of the Langevin equation then goes through in the same manner as previously described. The additional terms $\mathbbm{G}^{-1}W$ and $W\mathbbm{G}^{-1}$ become directly convoluted with $\Phi_1$. By construction, $\Phi_1$ has a suppressed short-wavelength component, and similarly to the discussion above, we expect the contribution of these terms to be negligible only for a sharp enough window function. 

Unlike in section \ref{sec:IReffective}, the inverse propagator expansion in \eqref{eq:InversePropExpansion} gives further contributions at higher order.
At next-to-next-to-leading order (NNLO) one finds
\beq
    \mathbbm{G}^{-1}\mathbbm{G}_\uv\mathbbm{G}^{-1}\mathbbm{G}_\uv\mathbbm{G}^{-1}&=\mathbbm{G}^{-1}{W}^2+{W}^2\mathbbm{G}^{-1}+{W}\mathbbm{G}^{-1}{W}-\mathbbm{G}^{-1}{W}^2\mathbbm{G}{W}\mathbbm{D}^{-1}\\
    &\quad -\mathbbm{G}^{-1}{W}\mathbbm{G}{W}^2\mathbbm{G}^{-1}-{W}\mathbbm{G}^{-1}{W}\mathbbm{G}{W}\mathbbm{G}^{-1}-\mathbbm{G}^{-1}{W}\mathbbm{G}{W}\mathbbm{G}^{-1}{W}\\
    &\quad + \mathbbm{G}^{-1}{W}\mathbbm{G}{W}\mathbbm{G}^{-1}+\mathbbm{G}^{-1}{W}\mathbbm{G}{W}\mathbbm{G}^{-1}{W}\mathbbm{G}{W}\mathbbm{G}^{-1},
    \label{NNLOInverseProp}
\eeq
where we note the reappearance of the NLO term  on the last line with the opposite sign. It is not manifest that \eqref{NNLOInverseProp} is subleading compared to \eqref{NLOInverseProp}, and we shall return to this point in the next section when considering a concrete example. 

\section{Testing approximations}
\label{sec:Testing}

In order to make explicit the central approximations needed to recover the Langevin equation in its standard form \eqref{eq:Langevin1} a smooth window function that separates the IR and UV scales is used as an example.
As discussed in the previous section, checking these approximations entails a comparison of the stochastic noise of \eqref{eq:noisecorr} with the white noise case \eqref{eq:StarobNoise2} and the justification to neglect the off-diagonal (c-q) components in \eqref{eq:offdiag}. In addition, we examine the expansion of the inverse IR propagator \eqref{eq:InversePropExpansion} from the alternative derivation presented in section \ref{sec:propagatordecomp}. Lastly, we study the impact of the choice of UV mode function on the noise correlations.
We take the common example of a Gaussian as a smooth window function, 
\beq
    W(k,t) = 1-e^{-\frac{1}{2}\left(\frac{k}{\sigma a_tH}\right)^2}.
    \label{eq:Gaussianwindow}
\eeq
The Gaussian window function fulfills the criteria set out in Ref.~\cite{Winitzki:1999ve}, and is parameterised by a dimensionless parameter $\sigma$, controlling both the location and steepness of the transition from IR to UV. This is in contrast to a smoothed step function (which will also be considered in section \ref{sec:noiseamplitude}), where the location is fixed by a parameter $\mu$ (as in (\ref{eq:stepfunction})) and $\sigma$ controls the steepness of the transition. In other words, the Gaussian does not become a step function in the limit $\sigma\rightarrow 0$, since the cutoff besides from steepening then also moves $k\rightarrow 0$. The Gaussian window function has been considered in a number of works, including \cite{Matarrese:2003ye,Liguori:2004fa}.

\subsection{Noise correlations}
\label{sec:Gaussian}

The noise correlation \eqref{eq:noisecorr} can be simplified slightly using that the time derivative of the full mode solution can be factorised as
\beq
    \dot{\phi}(k,t) = q_\nu(k,t)H\phi(k,t),
\eeq 
such that 
\beq
    \av{\xi(x)\xi(x')} =  \Re \int\frac{k\dd k}{2\pi^2 r}\sin kr\tilde{\mathcal{Q}}(k,t) \tilde{\mathcal{Q}}^*(k,t') \phi(k,t)\phi^*(k,t').
    \label{eq:noisecorr2}
\eeq
Here we have defined
\beq
    \tilde{\mathcal{Q}}(k,t) \equiv -\Ddot{W}(k,t)-[3+2 q_\nu(k,t)]H\dot{W}(k,t).
    \label{eq:newQ}
\eeq
where $q_\nu$ may be expressed as
\beq
    q_\nu(k,t)=-\epsilon_M-\frac{k}{a_tH}
    \frac{H_{\nu-1}^{(1)}\big(\frac{k}{a_tH(1-\epsilon_H)}\big)}{H_{\nu}^{(1)}\big(\frac{k}{a_tH(1-\epsilon_H)}\big)}\;.
    \label{eq:BigQ}
\eeq
to first order in $\epsilon_M,\epsilon_H$. We will proceed with $\epsilon_H=0$ and generalise to $\epsilon_H\neq 0$ later.

To begin with, using the approximate mode solution (\ref{eq:IRappmode}) for which $q_\nu(k,t)\simeq-\epsilon_M$, the noise correlator is found to be
\beq
    \av{\xi(x)\xi(x')} &= \frac{H^6}{8\pi^2}\sigma^{2\epsilon_M}(\sech H\tau)^{2+\epsilon_M}\Big[\sech^2 H\tau\Gamma[4+\epsilon_M]{}_1F_1\Big(4+\epsilon_M;\frac{3}{2};-\frac{r^2}{4\alpha}\Big)\\
    &\qquad +2(1-2\epsilon_M)\Gamma[3+\epsilon_M]{}_1F_1\Big(3+\epsilon_M;\frac{3}{2};-\frac{r^2}{4\alpha}\Big)\\
    &\qquad +(1-2\epsilon_M)^2\Gamma[2+\epsilon_M]{}_1F_1\Big(2+\epsilon_M;\frac{3}{2};-\frac{r^2}{4\alpha}\Big) \Big], 
    \label{eq:GaussianIRnoise}
\eeq
where $\tau\equiv t-t'$ denotes the time separation, $r\equiv |\mathbf{x}-\mathbf{x'}|$ the spatial separation, 
${}_1F_1(a,b,z)$ is a confluent hypergeometric function~\cite{grad}, and 
\beq
    \alpha \equiv \frac{1}{2\sigma^2H^2}\Big(\frac{1}{a_t^2}+\frac{1}{a_{t'}^2} \Big) = \frac{R^2}{\sigma^2\sech H\tau}\qquad \text{with}\qquad 
    R^2\equiv \frac{1}{a(t)a(t')H^2}\;.
    \label{eq:alpha}
\eeq

\begin{figure}
\centering
\begin{minipage}[t]{0.45\textwidth}
\hspace{-1cm}
    \begin{tikzpicture}
    \begin{axis}[width=1.1\textwidth,height=1.1\textwidth-2cm,
    scaled ticks=false,
    xmin=0,
    xmax=8,
    ymin=-2,
    ymax=6,
    ytick = {0,2,4},
    ylabel shift = -0.1cm,
    xlabel = { $H\tau$},
    ylabel = {$\frac{4\pi^2}{H^6}\av{\xi(x)\xi(x')}$},
    extra y ticks = {0},
    extra y tick labels={},
    extra tick style={grid=major},
    ]
    \addplot[blue] table [col sep=comma] {figures/IRapproxGaussian/IRGaussianVStauSigma1.csv};
    \addlegendentry{$\sigma=1.0$}
    \addplot[red] table [col sep=comma] {figures/IRapproxGaussian/IRGaussianVStauSigma05.csv};
    \addlegendentry{$\sigma=0.5$}
    \addplot[orange] table [col sep=comma] {figures/IRapproxGaussian/IRGaussianVStauSigma03.csv};
    \addlegendentry{$\sigma=0.3$}
    \end{axis}
    \end{tikzpicture}    
    \end{minipage}
    \begin{minipage}[t]{0.45\textwidth}
    \begin{tikzpicture}
    \begin{axis}[width=1.1\textwidth,height=1.1\textwidth-2cm,
    scaled ticks=false,
    xmin=0,
    xmax=10,
    ymin=-2,
    ymax=6,
    ytick = {0,2,4},
    ylabel shift = -0.1cm,
    xlabel = { $r/R_H$},
    ylabel = {$\frac{4\pi^2}{H^6}\av{\xi(x)\xi(x')}$},
    extra y ticks = {0},
    extra y tick labels={},
    extra tick style={grid=major},
    ]
    \addplot[blue] table [col sep=comma] {figures/IRapproxGaussian/IRGaussianVSrSigma1.csv};
    \addlegendentry{$\sigma=1.0$}
    \addplot[red] table [col sep=comma] {figures/IRapproxGaussian/IRGaussianVSrSigma05.csv};
    \addlegendentry{$\sigma=0.5$}
    \addplot[orange] table [col sep=comma] {figures/IRapproxGaussian/IRGaussianVSrSigma03.csv};
    \addlegendentry{$\sigma=0.3$}
    \end{axis}
    \end{tikzpicture}
    \end{minipage}
    \caption{Noise correlation \eqref{eq:GaussianIRnoise} with $\epsilon_M=0.001$ as a function of spatial separation $r/R_H$ with Hubble scale $R_H\equiv R(t=t')$ (left panel) and as a function of time separation $\tau\equiv t-t'$ for fixed spatial separation $r/R=1$ (right panel).}
    \label{fig:GaussianIRNoise}
\end{figure}
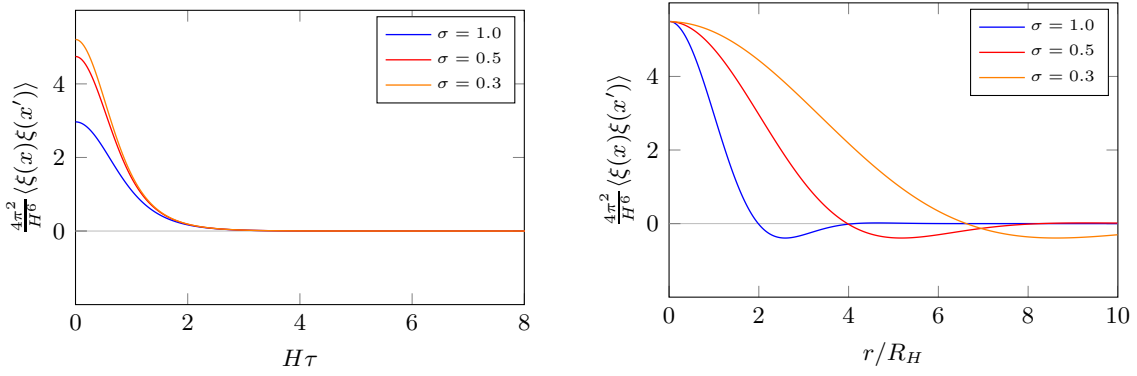
Because a number of similar expressions will appear later, it is worth pausing at this point. For equal time $t=t'$, we see that $R$ is the Hubble scale in time, and
the spatial separation $r$ then enters the correlation in units of the Gaussian width $(R/\sigma)^2$. Away from the equal-time limit, $r^2/4\alpha$ grows exponentially with $\tau$, resulting in the whole expression decaying exponentially (see left panel of Fig. \ref{fig:GaussianIRNoise}). 
For large spatial separations $r \gg R$, the noise correlation decays exponentially (see right panel of Fig. \ref{fig:GaussianIRNoise}), signifying that the noise evolution can be considered local for patches of length scales $r<R/\sigma$. It becomes apparent that for smaller $\sigma$, the decay is slower. On the other hand, since $R$ decreases very fast in time, the localisation in space is not very sensitive to the choice of $\sigma$.

In addition, due to the overall factor of $\sigma^{2\epsilon_M}$, the limit $\sigma\rightarrow 0$ does not commute with the massless limit. This factor can be replaced by 1 if we choose $\sigma$ in such a way that $\ln\sigma\ll \epsilon_M^{-1}/2$. If $\phi$ is the inflaton, in which case $\epsilon_M\simeq 10^{-5}-10^{-6}$, then this requirement is easily satisfied, however in principle one may tune $\sigma$ and $\epsilon_M$ to give any noise amplitude.

To get a better understanding of the contributions to the noise correlation \eqref{eq:GaussianIRnoise}, one may consider the decomposed noise terms $\xi_\phi$ and  $\xi_\pi$ of \eqref{eq:decomposedNoises}, where schematically
\beq
    \dot{\xi}_\phi(x) = \int\frac{\dd^3\mathbf{k}}{(2\pi)^\frac{3}{2}}\big(\ddot{W}_t+q_\nu(k,t)\dot{W}_t\big)\big[a_\mathbf{k}\phi(k,t)e^{i\mathbf{k}\cdot\mathbf{x}}+\text{h.c.}\big].
    \label{eq:dotXinoise}
\eeq
In particular, we obtain 
\beq
    \av{\xi_\phi(x)\xi_\phi(x')} &= \frac{H^4}{8\pi^2}\sigma^{2\epsilon_M}\Gamma[2+\epsilon_M](\sech H\tau)^{2+\epsilon_M}{}_1F_1\Big(2+\epsilon_M,\frac{3}{2},-\frac{r^2}{4\alpha} \Big)\\
    &\simeq \frac{H^4}{8\pi^2}\sech^2H\tau + \mathcal{O}(\epsilon_M),
    \label{eq:Gaussian_noise_phicorr}
\eeq
where the last equality applies for $r=0$.
The long-wavelength approximate result \eqref{eq:Gaussian_noise_phicorr} reproduces the result at leading order in $\sigma$ obtained using the exact massless mode solution \eqref{eq:dSmodesolution} in earlier works \cite{Winitzki:1999ve,Matarrese:2003ye}.

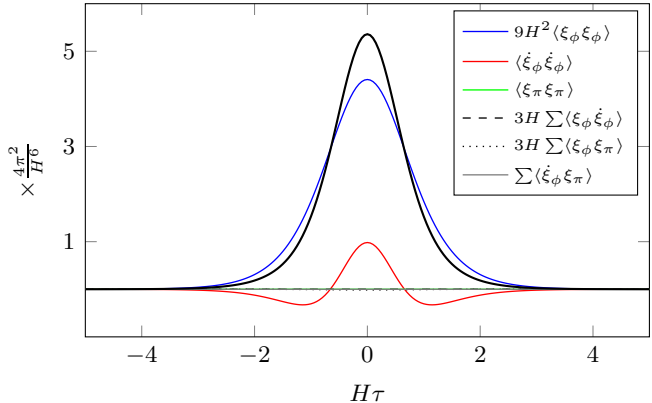
\begin{figure}
\centering
\begin{tikzpicture}
    \begin{axis}[width=9cm,height=6cm,
    xmin=-5,
    xmax=5,
    ymin=-1,
    ymax=6,
    ytick = {1,3,5},
    scaled ticks=false,
    xlabel = {$H\tau$},
    ylabel = {$\times\frac{4\pi^2}{H^6}$},
    scaled ticks=false,
    legend cell align={left},
    ]
    \addplot[blue] table [col sep=comma] {figures/IRapproxGaussian/xiPHIxiPHIdat.csv};
    \addlegendentry{$9H^2\av{\xi_\phi\xi_\phi}$}
    \addplot[red] table [col sep=comma] {figures/IRapproxGaussian/dotxiPHIdotxiPHIdat.csv};
    \addlegendentry{$\av{\dot{\xi}_\phi\dot{\xi}_\phi}$}
    \addplot[green] table [col sep=comma] {figures/IRapproxGaussian/xiPIxiPIdat.csv};
    \addlegendentry{$\av{\xi_\pi\xi_\pi}$}
    \addplot[dashed,black] table [col sep=comma] {figures/IRapproxGaussian/CrossxiPHIdotxiPHIdat.csv};
    \addlegendentry{$3H\sum\av{\xi_\phi\dot{\xi}_\phi}$}
    \addplot[dotted,black] table [col sep=comma] {figures/IRapproxGaussian/CrossxiPIxiPHIdat.csv};
    \addlegendentry{$3H\sum\av{\xi_\phi\xi_\pi}$}
    \addplot[gray] table [col sep=comma] {figures/IRapproxGaussian/CrossxiPIdotxiPHIdat.csv};
    \addlegendentry{$\sum\av{\dot{\xi}_\phi\xi_\pi}$}
    \addplot[black,thick] table [col sep=comma] {figures/IRapproxGaussian/FULLxixidat.csv};
    \end{axis}
\end{tikzpicture}
    \caption{Decomposed noise correlations at coincident spatial points for the Gaussian window with $\sigma=0.1$ and $\epsilon_M=0.005$. Here we denote $\sum\av{\xi_\phi\dot{\xi}_\phi}\equiv\av{\xi_\phi(x)\dot{\xi}_\phi(x')}+\av{\dot{\xi}_\phi(x)\xi_\phi(x')}$, etc. The combined noise $\av{\xi(x)\xi(x')}$ of \eqref{eq:GaussianIRnoise} is displayed in black for comparison.}
    \label{fig:GaussianIRAnalyticNoises}
\end{figure}

For the correlation of \eqref{eq:dotXinoise} one finds
\beq
    \av{\dot{\xi}_\phi(x)\dot{\xi}_\phi(x')} &= \frac{H^6}{8\pi^2}\sigma^{2\epsilon_M}(\sech H\tau)^{2+\epsilon_M}\Big[\Gamma[4+\epsilon_M]\sech^2H\tau{}_1F_1\Big(4+\epsilon_M,\frac{3}{2},-\frac{r^2}{4\alpha} \Big) \\
    &\quad-4\Gamma[3+\epsilon]{}_1F_1\Big(3+\epsilon_M,\frac{3}{2},-\frac{r^2}{4\alpha} \Big)+4\Gamma[2+\epsilon]{}_1F_1\Big(2+\epsilon_M,\frac{3}{2},-\frac{r^2}{4\alpha} \Big)\Big]\\
    &\simeq \frac{H^6}{8\pi^2}\sech^2H\tau(6\sech^2H\tau-4)+\mathcal{O}(\epsilon_M).
    \label{eq:gaussianddotWddotWcorrelator123}
\eeq
with the last equality again in the limit $r\rightarrow0$.
For the cross terms, we find   
\beq
    \av{\dot{\xi}_\phi(x){\xi}_\phi(x')}+\av{{\xi}_\phi(x) \dot{\xi}_\phi(x')} &= \frac{H^5}{4\pi^2}\sigma^{2\epsilon_M}(\sech H\tau)^{2+\epsilon_M}\Big[\Gamma[3+\epsilon_M] {}_1F_1\Big(3+\epsilon_M,\frac{3}{2},-\frac{r^2}{4\alpha} \Big) \\ 
    &\qquad -2\Gamma[2+\epsilon]{}_1F_1\Big(2+\epsilon_M,\frac{3}{2},-\frac{r^2}{4\alpha} \Big)\Big]\\
    &\simeq 0+\mathcal{O}(\epsilon_M).
    \label{eq:GaussiansumddotWdotWcorrelators}
\eeq
Correlations involving $\xi_\pi \propto q_\nu\simeq -\epsilon_M$ are  similarly suppressed. In Fig. \ref{fig:GaussianIRAnalyticNoises}, we show all the contributions to the noise correlators at $r=0$ as a function of time separation $H\tau$. We see that $\xi_\phi$ indeed dominates, while $\dot{\xi}_\phi$ has both positive and negative contributions. 

The noise correlator is localised around $H\tau=0$, but has more features than simply a delta-function. Nevertheless, a final connection to the white noise delta function can be made, even for finite $\sigma$, through the replacement (for this example, in the massless limit $\epsilon_M=0$)
\beq
\sech^2H\tau \rightarrow 2\delta(H\tau)= \frac{2}{H}\delta(t-t'), 
\label{eq:deltareplacement}
\eeq
where the factor of 2 is chosen so that
\beq
\int \dd (H\tau) \sech^2H\tau = \frac{2}{H}\int \dd\tau \delta(t-t').
\label{eq:timelocalization}
\eeq
In other words, we replace a moderately localised noise distribution by a fully localised one with the same integrated power, i.e., such that
\beq
    \av{\xi_\phi(x)\xi_\phi(x')} \simeq \frac{H^4}{8\pi^2}\sech^2H\tau \simeq \frac{H^3}{4\pi^2}\delta(t-t').
    \label{eq:GaussianNoise3}
\eeq
With this procedure, it can be seen that the integral of the $\av{\dot{\xi}_\phi\dot{\xi}_\phi}$ correlator vanishes identically to leading order, and so non-local time correlations (positive and negative) are ignored in this procedure. Only the $\av{\xi_\phi\xi_\phi}$ contribution remains.

In section \ref{sec:noiseamplitude}, we will further investigate how the noise distribution depends on the choice of window function and its parameters, as well as 
the choice of approximation for the UV mode functions (\ref{eq:SRdSmodesolution})--(\ref{eq:IRappmode}). The dependence on the mass parameter $\epsilon_M$ and the generalisation beyond leading order
in slow-roll, $\epsilon_H\neq 0$, will also be considered. Ultimately, we will compute the amplitude of the localised noise, making use of the prescription (\ref{eq:deltareplacement}). The prefactor of the delta-function will be referred to as the noise amplitude. 

Before doing so, however, we will use the Gaussian window function to explicitly check some of the assumptions made with the derivations in the previous section. The conclusions will hold also for the other window functions considered below.

\subsection{Off-diagonal (q-c) contributions to the IR field evolution}
\label{sec:checkingGR}
In our derivation of the stochastic evolution equation for the IR modes (\ref{eq:LangevinEoM}), we made the claim that the off-diagonal c-q contributions, including the term $\Upsilon$, can be neglected. In the following we will check this explicitly for the terms that do not contain any operators acting directly on the IR field $\phi_\ir$.

With the approximate solution \eqref{eq:IRappmode} for the dS UV mode functions, we first find for the retarded propagator 
\beq 
    G_R(k,t,t') \simeq -\frac{H^2}{3}\Theta(t-t')\left[(a_tH)^{-3+\epsilon_M}(a_{t'}H)^{-\epsilon_M}-(a_{t'}H)^{-3+\epsilon_M}(a_{t}H)^{-\epsilon_M}\right],
\eeq
which for $\epsilon_M\rightarrow 0$ goes as $a^{-3}(t)$.
We can then straightforwardly compute
\beq
    \Upsilon(x,x') &= \frac{H^7}{4\pi^2}\sigma^3(a_ta_{t'})^{\frac{3}{2}}\sech^\frac{7}{2} H\tau\Big[\Gamma\left[\tfrac{11}{2}\right]\sech^2 H\tau{}_1F_1\Big(\frac{11}{2},\frac{3}{2},-\frac{r^2}{4\alpha}\Big)\\
    &\qquad+2\Gamma\left[\tfrac{9}{2}\right]{}_1F_1\Big(\frac{9}{2},\frac{3}{2},-\frac{r^2}{4\alpha}\Big)\Big(1+\sech H\tau\Big(\frac{a_{t'}}{a_t}\frac{\partial_{t'}}{H}+\frac{a_{t}}{a_{t'}}\frac{\partial_{t}}{H}\Big)
    \Big) \\
    &\qquad+(-5+12\epsilon_M)\Gamma\left[\tfrac{7}{2}\right]{}_1F_1\Big(\frac{7}{2},\frac{3}{2},-\frac{r^2}{4\alpha}\Big)\Big]G_R(x,x')\\
    &\qquad -\frac{H^6}{3\pi^2}(3-2\epsilon_M)\sigma^3 \Gamma\left[\tfrac{7}{2}\right] {}_1F_1\Big(\frac{7}{2},\frac{3}{2},-\frac{1}{4}(r\sigma a_tH)^2\Big)\delta (H\tau),
\eeq
where we have used
\beq
(\partial_t+\partial_{t'})G_R= -3HG_R,\quad \partial_t\partial_{t'}G_R = (3-\epsilon_M)\epsilon_M H^2 G_R + \frac{\delta(t-t')}{\Theta(t-t')}\partial_{t'}G_R.
\eeq
As an example, in the $r=0$, $\epsilon_M=0$ limit this reduces to
\beq
\Upsilon(x,x') &= \frac{H^6}{\pi^2}\sigma^3\Gamma\left[\tfrac{7}{2}\right]\Big[\sech^{\tfrac{7}{2}}H\tau
    \Big(\tfrac{21}{8}\sinh\tfrac{3}{2}H\tau\sech^2H\tau \\
    &\qquad-\tfrac{7}{2}\sinh\tfrac{1}{2}H\tau\sech H\tau
    +\tfrac{1}{3}\sinh\tfrac{3}{2}H\tau \Big)\Theta(\tau)-\delta(H\tau)\Big].
\eeq
The important result is a suppression by an overall factor of $\sigma^3$, but not by e.g. powers of the scale factor. This is still much faster than for the diagonal correlator \eqref{eq:GaussianIRnoise}, which has a leading $\sigma^{2\epsilon_M}$.

Similarly, in \eqref{eq:offdiag}, one of the terms may by partial integration be made not to act
explicitly on the $\phi_\ir$. We find for this term,
\beq
    \frac{2}{a_{t'}^3}&\delta(t-t')\int\frac{\dd^3\mathbf{k}}{(2\pi)^3}e^{i\mathbf{k}\cdot(\mathbf{x}-\mathbf{x'})}\tilde{Q}_{t'}{W}_{t'} \\
    &= \delta(H\tau)\frac{H^6\sigma^3}{2\pi^2}\Big[3\Gamma[\tfrac{7}{2}]\left({}_1F_1(\tfrac{7}{2},\tfrac{3}{2},-\tfrac{1}{4}(r\sigma aH)^2)-2^{\frac{7}{2}} {}_1F_1(\tfrac{7}{2},\tfrac{3}{2},-\tfrac{1}{2}(r\sigma aH)^2) \right)\\
    &\qquad+\Gamma\left[\tfrac{5}{2}\right]\left({}_1F_1(\tfrac{5}{2},\tfrac{3}{2},-\tfrac{1}{4}(r\sigma aH)^2)-2^{\frac{5}{2}}{}_1F_1(\tfrac{5}{2},\tfrac{3}{2},-\tfrac{1}{2}(r\sigma aH)^2) \right)\Big]\\&\simeq \frac{H^6}{\pi^2}\sigma^3\Gamma\left[\tfrac{7}{2}\right]\bigg(\frac{17-128\sqrt{2}}{10}\bigg)\delta(H\tau),
\eeq
where again the limits $r=0$, $\epsilon_M=0$ are taken in the last line. It is clear that also this quantity is suppressed by $\sigma^3$.

In Ref.~\cite{Morikawa1990}, $\Upsilon$ was also reported to decay as $\mu^3$ for a step window function \eqref{eq:stepfunction}. This is consistent, since for the Gaussian window function, $\sigma$ also parametrises the cutoff $\mu$. We conclude that neglecting the off-diagonal contributions in the influence functional is consistent in the small-$\sigma$ (or $\mu$) regime, for $\sigma^3\ll\sigma^{2\epsilon_M}$.

\subsection{Inverse IR propagator expansion}
\label{sec:propagatorexpansion}

\begin{figure}
\centering
\begin{tikzpicture}
    \begin{axis}[width=9cm,height=6cm,
    xmin=0,
    xmax=0.98,
    ymin=0.7,
    ymax=1.1,
    xtick = {0.2,0.4,0.6,0.8},
    ytick = {0.7,0.8,0.9,1.0},
    scaled ticks=false,
    ylabel shift = -0.1cm,
    xlabel = {$\sigma$},
    ylabel = {$\mathcal{N}_\text{NNLO}/\mathcal{N}_\text{NLO}$},
    scaled ticks=false,
    legend pos=south west,
    ]
    \addplot[blue] table [col sep=comma] {"figures/GaussianNNLO_NLOratio/GaussianNNLO_NLOratio.csv"};
    \addlegendentry{$r/R=1.0$}
    \addplot[orange] table [col sep=comma] {"figures/GaussianNNLO_NLOratio/GaussianNNLO_NLOratio2.csv"};
    \addlegendentry{$r/R=0.1$}
    \end{axis}
\end{tikzpicture}
    \caption{The ratio of the noise amplitude from the NNLO contributions \eqref{leadingNNLOTerms} to the NLO noise \eqref{eq:GaussianIRnoise} (with the Gaussian window function) as a function of $\sigma$ with $\epsilon_M,\epsilon_H=0$.}
    \label{fig_GaussianNNLO_NLOratio}
\end{figure}
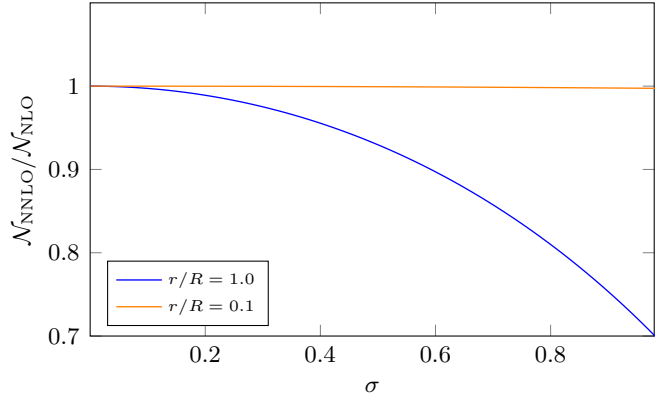

The expansion of the IR propagator \eqref{eq:InversePropExpansion} relies on the smallness
of the overlap between the IR field and the UV propagator. To at least get an idea of the convergence, we proceed to consider the stochastic noise at NNLO in \eqref{NNLOInverseProp}, given by the q-q component
\beq
\begin{aligned}
    i(\mathbbm{G}^{-1}\mathbbm{G}_\uv\mathbbm{G}^{-1}\mathbbm{G}_\uv\mathbbm{G}^{-1})_\text{qq} &= G^{-1}WG_FWG^{-1}-WG^{-1}WG_FWG^{-1}-G^{-1}WG_FWG^{-1}W\\
    &\quad-G^{-1}W^2G_FWG^{-1}-G^{-1}WG_FW^2G^{-1}\\
    &\quad+G^{-1}WG_FWG^{-1}WG_AWG^{-1}\\
    &\quad+G^{-1}WG_RWG^{-1}WG_FWG^{-1}.
    \label{StochasticNoiseNNLO}
\end{aligned}
\eeq
Assuming that it is valid to discard terms where the window function $W$ is directly convoluted with the IR field constituents, and in addition that the contribution involving retarded/advanced propagators is negligible, \eqref{StochasticNoiseNNLO} reduces to
\beq
    i(\mathbbm{G}^{-1}\mathbbm{G}_\uv\mathbbm{G}^{-1}\mathbbm{G}_\uv\mathbbm{G}^{-1})_\text{qq} \simeq G^{-1}WG_FWG^{-1}-G^{-1}W^2G_FWG^{-1}-G^{-1}WG_FW^2G^{-1}.
    \label{leadingNNLOTerms}
\eeq
Here we note that if the window function is a projection so that $W^2=W$, all the terms add up to become equal to the NLO contribution considered above \eqref{NLOInverseProp}. Clearly, the step function is a pathological case, for which this expansion fails.

For a general window function, one can compute the three contributions in \eqref{leadingNNLOTerms} explicitly. Using again the Gaussian window function as a test case, we obtain
\beq
\begin{aligned}
    G^{-1}W^2G_FWG^{-1} &= 2G^{-1}WG_FWG^{-1}-\frac{H^6}{4\pi^2}\sigma^{2\epsilon_M}\Big(\frac{2}{2e^{-H\tau}+e^{H\tau}}\Big)^{2+\epsilon_M}\\
    &\quad\times\Big[2\Gamma[4+\epsilon_M]\Big(\frac{2}{2e^{-H\tau}+e^{H\tau}}\Big)^{2}
    {}_1F_1\left(4+\epsilon_M,\frac{3}{2},-\frac{r^2}{4\tilde{\alpha}_t}\right)\\
    &\qquad+2(1-2\epsilon_M)\Gamma[3+\epsilon_M]{}_1F_1\Big(3+\epsilon_M,\frac{3}{2},-\frac{r^2}{4\tilde{\alpha}_t}\Big)\\
    &\qquad+(1-2\epsilon_M)^2\Gamma[2+\epsilon_M]{}_1F_1\Big(2+\epsilon_M,\frac{3}{2},-\frac{r^2}{4\tilde{\alpha}_t}\Big)\Big],
\end{aligned}\label{eq:NNLOnoise}
\eeq
with 
\beq 
    \tilde{\alpha}_t&\equiv\frac{2e^{-H\tau}+e^{H\tau}}{2\sigma^2a_ta_{t'}H^2},
\eeq
and similarly for $G^{-1}WGW^2G^{-1}$ with $2e^{-H\tau}+e^{H\tau} \leftrightarrow e^{-H\tau}+2e^{H\tau}$. We notice the reappearance of the NLO noise kernel $G^{-1}WG_FWG^{-1}$ as the first term on the right-hand-side of \eqref{eq:NNLOnoise}. 

To quantify the relation between the two orders of the expansion, we perform the localisation procedure of (\ref{eq:timelocalization}), i.e., replacing the noise by a delta function and matching the normalisation. The ratio of the noise amplitude for the NNLO noise terms \eqref{leadingNNLOTerms} and the NLO result \eqref{eq:GaussianIRnoise} is plotted in Fig. \ref{fig_GaussianNNLO_NLOratio}.  
Judging only by the subset of terms computed, there is no suppression of the NNLO contributions compared to that of NLO. Hence, we cannot confirm that truncating the inverse propagator expansion at NLO is a controlled approximation. 

\subsection{Exact or long-wavelength UV mode functions}
\label{sec:gaussmode}

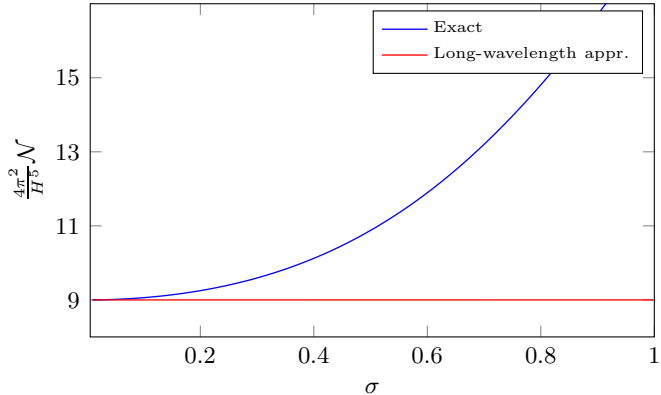
\begin{figure}
\centering
    \begin{tikzpicture}
    \hspace{-0.6cm}
    \begin{axis}[width=9cm,height=6cm,
    xmin=0.006,
    xmax=1,
    ymin=8,
    ymax=17,
    ytick ={9,11,13,15},
    ylabel shift = -0.1cm,
    scaled ticks=false,
    xlabel = {$\sigma$},
    ylabel = {$\frac{4\pi^2}{H^5}\mathcal{N}$},
    scaled ticks=false,
    legend cell align={left},
    ]
    \addplot[blue] table [col sep=comma] {figures/sigmadepGaussian/GaussianNoiseNormalisation.csv};
    \addlegendentry{Exact}
    \addplot[red] table [col sep=comma] {figures/sigmadepGaussian/GaussianNoiseNormalisationIRapprox.csv};
    \addlegendentry{Long-wavelength appr.}
    \end{axis}
    \end{tikzpicture}
    \caption{Noise amplitude using the Gaussian window function in massless dS using the exact \eqref{eq:dSmodesolution} and long-wavelength approximate  propagator \eqref{eq:IRappmode}
    as a function of $\sigma$ with sub-horizon spatial separation $r/R=0.01$. }
    \label{fig_GNoiseNormAppVSexact}
\end{figure}

Another approximation worthy of examination is the use of 
the IR-approximate mode solutions (\ref{eq:IRappmode}) instead of the exact mode solutions (\ref{eq:SRdSmodesolution}). The latter are complicated to treat analytically and so we compare the two numerically. As a first example, we continue our analysis of the Gaussian window function by computing the noise amplitude $\mathcal{N}$ with both the approximate and the exact mode functions, shown in Fig. \ref{fig_GNoiseNormAppVSexact}. For very small $\sigma$, the two can be seen to agree asymptotically, but diverge for larger values. Apparently, the $\sigma\rightarrow 0$ is a crucial ingredient of the standard result for the noise amplitude. 

\section{The noise amplitude and its parameter dependence}
\label{sec:noiseamplitude}

As we illustrated for the Gaussian window function, for spatial separations much smaller than the horizon, the noise correlations are localised near $t=t'$. For many smooth window functions, this is a general property, although the noise is not white in a strict sense. Making the interpretation of the distribution being a smoothed delta function, we can attempt to extract a characteristic noise amplitude as described in (\ref{eq:timelocalization}), suitable for the stochastic inflationary dynamics. Ultimately, this normalisation is what enters
into observables, as described in section \ref{sec:langevin}. To this end we consider the noise on patches larger than the cutoff scale and define the amplitude $\mathcal{N}$ as 
\beq
    \av{\xi(x)\xi(x')} \rightarrow \mathcal{N}\delta(t-t')\Theta(1-\mu raH),
    \label{eq:amplitude}
\eeq
where $\mathcal{N}$ is defined as described in section \ref{sec:Gaussian} through the matching
\beq
    \mathcal{N}\equiv\int \dd\tau \av{\xi\xi}(\tau).
\eeq
As mentioned in section \ref{sec:IRstochastic}, when the noise component $\xi_\phi$ dominates, we may write 
\beq
    \av{\xi(x)\xi(x')} \rightarrow 9H^2\delta(t-t')\Theta(1-\mu raH)\int \dd\tau \av{\xi_\phi\xi_\phi}(\tau).
    \label{eq:amplitudephiphi}
\eeq

In addition to the Gaussian window function, we will consider three smoothed step functions, all with the property than in certain limits, they become a strict step function (in contrast to the Gaussian). In particular we consider an error (Erfc) function, which turns out to be particularly tractable analytically. In section \ref{sec:otherwindows} we consider two other sigmoid functions for comparison, namely the Tanh and Arctan functions.

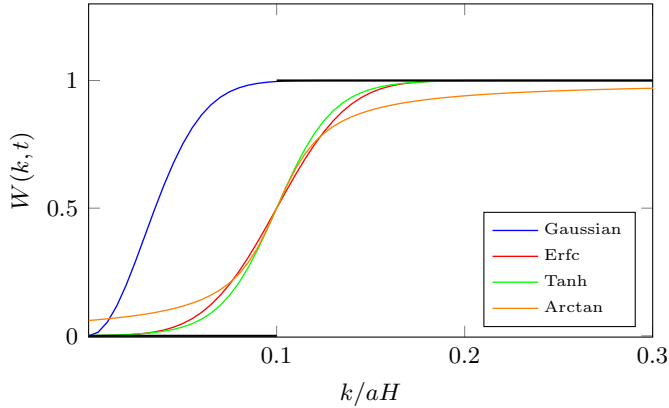
\begin{figure}
\centering
\begin{tikzpicture}
    \hspace{-0.5cm}
    \begin{axis}[width=9cm,height=6cm,
    xmin=0,
    xmax=0.3,
    ymin=-0.005,
    ymax=1.3,
    scaled ticks=false,
    xtick = {0.1,0.2,0.3},
    ylabel shift = -0.1cm,
    xlabel = {$k/aH$},
    ylabel = {$W(k,t)$},
    scaled ticks=false,
    legend pos=south east,
    legend cell align={left},
    ],
    \addplot[blue] table [col sep=comma] {figures/window/gaussdat.csv};
    \addlegendentry{Gaussian},
    \addplot[red] table [col sep=comma] {figures/window/erfcdat.csv};
    \addlegendentry{Erfc},
    \addplot[green] table [col sep=comma] {figures/window/tanhdat.csv};
    \addlegendentry{Tanh},
    \addplot[orange] table [col sep=comma] {figures/window/arctandat.csv};
    \addlegendentry{Arctan},
    \addplot[thick,black,domain=0:0.1] {0};
    \addplot[thick,black,domain=0.1:0.5] {1};
    \end{axis}
\end{tikzpicture}
\caption{Window functions with $\mu=0.1$ and $\sigma=0.03$ compared to a step function (black line). 
For the Gaussian, $\sigma$ parameterises both the cutoff scale and the sharpness of the transition.}
\label{fig:windows}
\end{figure}

For the smoothed step functions, the parameter $\mu$ controls the position of the cutoff, just as for the strict step function \eqref{eq:stepfunction}. Small $\mu$ means that the cutoff is far in the super-horizon IR region. The parameter $\sigma$ controls the width of the smoothed step function, and in the limit $\sigma\rightarrow 0$, we recover the sharp step function. This is illustrated in Fig. \ref{fig:windows}.

\subsection{Erfc window function}
\label{sec:errorfunction}

A smooth approximate step function can be constructed from the (complementary) error function as
\beq
    {W}(k,t) = \frac{1}{2}\mathrm{erfc}\left[\frac{1}{\sqrt{2}\sigma}\left(\mu-\frac{k}{a_tH}\right)\right].
\eeq
Inserting first the long-wavelength approximate mode solution \eqref{eq:IRappmode} into \eqref{eq:noisecorr2}, the noise correlation is found \cite{grad} to be
\beq
    \av{\xi(x)\xi(x')}&=-\frac{iH^6e^{M_1-\frac{\mu^2}{\sigma^2}}}{16\pi^3}\frac{R}{r}\bigg[\frac{1}{\sigma^6}\bigg(\mathcal{T}_{-(5+2\epsilon_M)}-\frac{2\mu}{\sech\frac{H\tau}{2}}\mathcal{T}_{-(4+2\epsilon_M)}+\mu^2\mathcal{T}_{-(3+2\epsilon_M)}\bigg)\\
    &\qquad +\frac{4(1-\epsilon_M)}{\sigma^4}\bigg(\frac{1}{\sech H\tau }\mathcal{T}_{-(3+2\epsilon_M)} -\frac{\mu}{\sech\frac{H\tau}{2}}\mathcal{T}_{-(2+2\epsilon_M)} \bigg)\\
    &\qquad+\frac{4(1-\epsilon_M)^2}{\sigma^2}\mathcal{T}_{-(1+2\epsilon_M)} \bigg].
    \label{eq_ErfcNoiseIRapproxAnalytical}
\eeq
Here we have defined 
\beq
    \mathcal{T}_{-n}\equiv \Gamma[n]\left(\frac{\sigma^2\sech H\tau}{2}\right)^{\frac{n}{2}} \left[\exp{M_2}D_{-n}(M_3)-\exp{M_2^*}D_{-n}(M_3^*) \right],
\eeq
where $D_{-n}(x)$ is the parabolic cylinder function. The arguments are
\begin{align}
    M_1 \equiv \frac{\beta^2-r^2}{8\alpha}\,,\qquad
    M_2 \equiv -\frac{ir\beta}{4\alpha}\,,\qquad
    M_3 \equiv \frac{\beta-ir}{\sqrt{2\alpha}}\,,
    \label{eq_Marguments}
\end{align}
with $\alpha$ as in \eqref{eq:alpha} and $\beta$ defined as 
\beq 
    R^2\equiv \frac{1}{a(t)a(t')H^2}\,,\qquad
    \alpha \equiv  \frac{R^2}{\sigma^2\sech H\tau}\,,\qquad 
    \beta \equiv -\frac{2\mu R}{\sigma^2\sech \frac{H\tau}{2}}\,.
\eeq
The cutoff scale $\mu$ is intended to be finite, in order to retain some modes in the IR. The sharpness parameter $\sigma$ can however be arbitrarily small. 
We are interested in the noise amplitude in the limit of a steep step
momentum transition $\sigma\ll 1$.
To study this limit, we proceed by eliminating the spatial dependence by expanding the noise \eqref{eq_ErfcNoiseIRapproxAnalytical} to leading order in spatial separation $\mu r/R\ll 1$,
\beq
    \av{\xi(x)\xi(x')}&\simeq \frac{H^6}{8\pi^3}e^{\frac{\mu^2}{\sigma^2}\left(\frac{1}{2}\frac{\sech H\tau}{\sech^2\frac{H\tau}{2}}-1\right)}\bigg[\frac{1}{\sigma^6}\bigg(\tilde{\mathcal{T}}_{-(5+2\epsilon_M)}
    -\frac{2\mu}{\sech\frac{H\tau}{2}}\tilde{\mathcal{T}}_{-(4+2\epsilon_M)}+\mu^2\mathcal{T}_{-(3+2\epsilon_M)}\bigg)\\
    &\qquad +\frac{4(1-\epsilon_M)}{\sigma^4}\bigg(\frac{1}{\sech H\tau }\tilde{\mathcal{T}}_{-(3+2\epsilon_M)} 
    -\frac{\mu}{\sech\frac{H\tau}{2}}\tilde{\mathcal{T}}_{-(2+2\epsilon_M)} \bigg)\\
    &\qquad+\frac{4(1-\epsilon_M)^2}{\sigma^2}\tilde{\mathcal{T}}_{-(1+2\epsilon_M)} \bigg],
    \label{eq:smallParamNoise}
\eeq
where\footnote{We use the recurrence relation $D_{-n}(x)-\frac{1}{x}D_{-n+1}(x)=-\frac{n}{x}D_{-n-1}(x)$.}
\beq
    \tilde{\mathcal{T}}_{-n}\equiv \Gamma[n+1]\bigg(\frac{\sigma^2\sech H\tau}{2}\bigg)^{\frac{n+1}{2}} D_{-n-1}\bigg(-\sqrt{2}\frac{\mu}{\sigma}\frac{\sqrt{\sech H\tau}}{\sech\frac{H\tau}{2}}\bigg).
    \label{eq:TildeT1}
\eeq
For large negative arguments, this may be further approximated as\footnote{The parabolic cylinder functions have the asymptotic behaviour 
$$D_{-n-1}(-|x|)\big|_{x\rightarrow\infty}\simeq  \frac{\sqrt{2\pi}}{\Gamma[n+1]}e^{x^2/4}|x|^n\left[1+\frac{n(n-1)}{2x^2} +\frac{n(n-1)(n-2)(n-3)}{8x^4}+\dotso\right].$$}
\beq
    \tilde{\mathcal{T}}_{-n}&\simeq \sqrt{\pi}e^{\frac{1}{2}\frac{\mu^2}{\sigma^2}\frac{\sech H\tau}{\sech^2\frac{H\tau}{2}}}
    \left(\mu\frac{\sech H\tau}{\sech\frac{H\tau}{2}}\right)^n \sigma\sqrt{\sech H\tau}\bigg[1+\left(\frac{\sigma}{\mu}\right)^2\frac{n(n-1)}{4}\frac{\sech^2\frac{H\tau}{2}}{\sech H\tau}\\
    &\qquad+\left(\frac{\sigma}{\mu}\right)^4\frac{n(n-1)(n-2)(n-3)}{32}\frac{\sech^4\frac{H\tau}{2}}{\sech^2 H\tau}+\dotso\bigg], 
    \label{eq:tildeTapprox}
\eeq
which is then valid for large values of $\mu/\sigma$.
In the limit $\epsilon_M\rightarrow0$, the noise correlator becomes
\beq 
    \av{\xi(x)\xi(x')} &\simeq \frac{H^6}{4\pi^2} \frac{1}{2\sqrt{\pi}}e^{\frac{\mu^2}{\sigma^2}\left(\frac{\sech H\tau}{\sech\frac{H\tau}{2}}-1\right)}\bigg[-\left(\frac{\mu}{\sigma}\right)^5\cosh^3\tfrac{H\tau}{2}\sech^\frac{11}{2} H\tau\sinh^2\tfrac{H\tau}{2}\\
    &\qquad  +\frac{1}{4}\left(\frac{\mu}{\sigma}\right)^3 \cosh \tfrac{H\tau}{2} (7-2 \cosh H\tau -3 \cosh2H\tau) \sech^{\frac{9}{2}}H\tau\\
    &\qquad + \frac{1}{4} \frac{\mu}{\sigma}\left(31-6\cosh H\tau +16 \cosh2H\tau\right)\cosh\tfrac{H\tau}{2} \sech^{\frac{7}{2}}H\tau\bigg].
    \label{eq:noiseExpanded}
\eeq

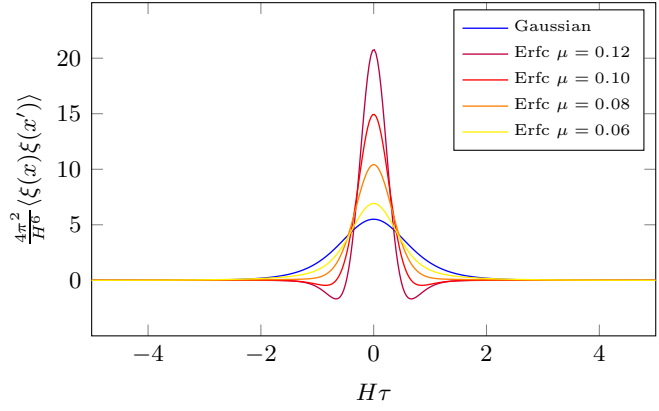
\begin{figure}[t]
\centering
\begin{tikzpicture}
    \begin{axis}[width=9cm,height=6cm,
    xmin=-5,
    xmax=5,
    ymin=-5,
    ymax=25,
    ytick = {0,5,10,15,20},
    scaled ticks=false,
    xlabel = {$H\tau$},
    ylabel = {$\frac{4\pi^2}{H^6}\av{\xi(x)\xi(x')}$},
    ylabel shift = -0.1cm,
    scaled ticks=false,
    extra tick style={grid=major},
    legend cell align={left},
    ]
    \addplot[blue] table [col sep=comma] {figures/timesepdep/GaussianNoiseVSTimeSepData.csv};
    \addlegendentry{Gaussian}
    \addplot[purple] table [col sep=comma] {figures/timesepdep/ErfcNoiseVSTimeSep012.csv};
    \addlegendentry{Erfc $\mu=0.12$}
    \addplot[red] table [col sep=comma] {figures/timesepdep/ErfcNoiseVSTimeSep010.csv};
    \addlegendentry{Erfc $\mu=0.10$}
    \addplot[orange] table [col sep=comma] {figures/timesepdep/ErfcNoiseVSTimeSep008.csv};
    \addlegendentry{Erfc $\mu=0.08$}
    \addplot[yellow] table [col sep=comma] {figures/timesepdep/ErfcNoiseVSTimeSep006.csv};
    \addlegendentry{Erfc $\mu=0.06$}
    \end{axis}
\end{tikzpicture}
\caption{Noise correlation as a function of time separation $\tau\equiv t-t'$ for the Erfc and Gaussian window functions with $\sigma=0.03$, $\epsilon_M=0$ and $r/R=0.01$.}
\label{fig:erfctaudep}
\end{figure}

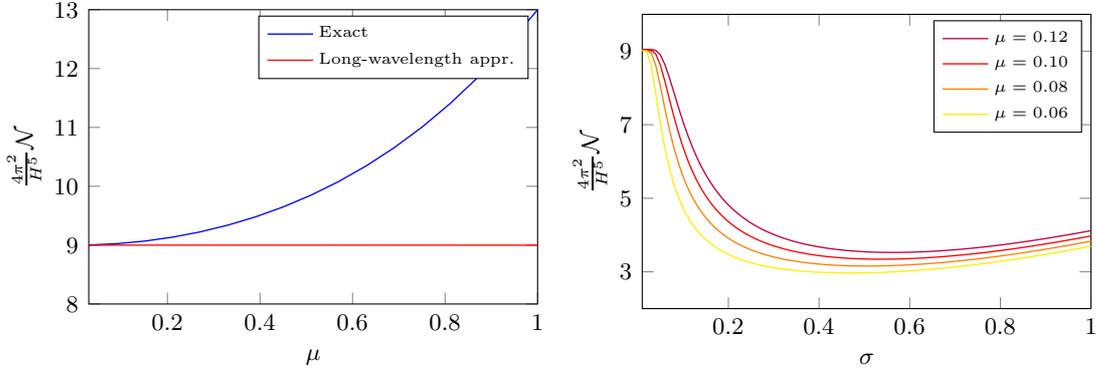
\begin{figure}[t]
\centering
\begin{subfigure}[t]{0.45\textwidth}
\begin{tikzpicture}
    \begin{axis}[width=1.1\textwidth,height=1.1\textwidth-2cm,
    scaled ticks=false,
    xmin=0.03,
    xmax=1,
    ymin=8,
    ymax=13,
    ytick = {8,9,10,11,12,13},
    xlabel = { $\mu$},
    ylabel = {$\frac{4\pi^2}{H^5}\mathcal{N}$},
    legend cell align={left},
    ylabel shift = -0.1cm,
    ]
    \addplot[blue] table [col sep=comma] {figures/mudepErfc/ErfcNormVSmuSigma0005data.csv};
    \addlegendentry{Exact}
    \addplot[red] table [col sep=comma] {figures/mudepErfc/ErfcNormVSmuIRapp.csv};
    \addlegendentry{Long-wavelength appr.}
    \end{axis}
    \end{tikzpicture}
\end{subfigure}
\hfill
\begin{subfigure}[t]{0.45\textwidth}
\hspace{-1cm}
\begin{tikzpicture}
    \begin{axis}[width=1.1\textwidth,height=1.1\textwidth-2cm,
    xmin=0.01,
    xmax=1,
    ymin=2,
    ymax=10,
    ytick = {3,5,7,9},
    scaled ticks=false,
    xlabel = {$\sigma$},
    ylabel = {$\frac{4\pi^2}{H^5}\mathcal{N}$},
    ylabel shift = -0.1cm,
    scaled ticks=false,
    ]
    \addplot[purple] table [col sep=comma] {figures/sigmadepErfc/ErfcNoiseNormMu012.csv};
    \addlegendentry{$\mu=0.12$}
    \addplot[red] table [col sep=comma] {figures/sigmadepErfc/ErfcNoiseNormMu010.csv};
    \addlegendentry{$\mu=0.10$}
    \addplot[orange] table [col sep=comma] {figures/sigmadepErfc/ErfcNoiseNormMu008.csv};
    \addlegendentry{$\mu=0.08$}
    \addplot[yellow] table [col sep=comma] {figures/sigmadepErfc/ErfcNoiseNormMu006.csv};
    \addlegendentry{$\mu=0.06$}
    \end{axis}
\end{tikzpicture}
\end{subfigure}
    \caption{Left panel: Noise amplitudes for the Erfc window function with exact \eqref{eq:dSmodesolution} and long-wavelength approximate \eqref{eq:IRappmode} massless dS propagators as a function of $\mu$, with $\sigma=0.005$ and spatial separation set to $r/R=0.01$. Right panel: Noise amplitudes for different $\mu$ computed with the exact propagator \eqref{eq:dSmodesolution} as a function of $\sigma$.}
    \label{fig_Erfc_IntegratedNoiseVSsigmaExact}
\end{figure}

\paragraph{Noise localisation and amplitude.}
The noise distribution \eqref{eq:noiseExpanded} is centered around $t=t'$, and becomes more localised for increasing values of the ratio $\mu/\sigma$, as illustrated in Fig. \ref{fig:erfctaudep}. Therefore we may expand in $H\tau$, to find
\beq
    \av{\xi(x)\xi(x')}&\simeq \frac{H^6}{4\pi^2}\frac{1}{2\sqrt{\pi}} e^{\left(\frac{\mu}{\sigma}\right)^2\left(-\frac{1}{4}(H\tau)^2+\frac{5}{48}(H\tau)^4-\frac{61}{1440}(H\tau)^6 +\dotso\right)} \bigg[\frac{\sigma}{\mu}\left(\frac{9}{4}+8\right)\\
    &\qquad+ \left(\frac{\sigma}{\mu}\right)^3\left[\frac{1}{2}-\frac{45}{16}(H\tau)^2\right]+ \left(\frac{\sigma}{\mu}\right)^5\left[-\frac{1}{4}(H\tau)^2 +\frac{55}{96}(H\tau)^4\right]+ \dotso\bigg].
    \label{eq:noiseExpanded2}
\eeq
The noise amplitude is then obtained by integrating \eqref{eq:noiseExpanded2} over time separation. 
It turns out that some contributions are recovered keeping only the first-order term in the exponential, i.e., following a Gaussian distribution,
\beq
    &\frac{H^6}{4\pi^2}\frac{1}{2\sqrt{\pi}}\int \dd (H\tau) \bigg[ \bigg(\frac{9}{4}+8\bigg)\frac{\mu}{\sigma}-\frac{45}{16}\left(\frac{\sigma}{\mu}\right)^3(H\tau)^2+\frac{55}{96}\left(\frac{\sigma}{\mu}\right)^5(H\tau)^4\bigg]e^{-\frac{1}{4}\left(\frac{\mu}{\sigma}\right)^2(H\tau)^2} \\
    &= \bigg(\frac{7}{2}+8\bigg)\frac{H^6}{4\pi^2},
\eeq
while two terms receive corrections from the higher-order terms in the exponential, 
\beq 
    \frac{H^6}{4\pi^2}\frac{1}{2\sqrt{\pi}}\int \dd (H\tau) \left[-\frac{1}{4}\left(\frac{\sigma}{\mu}\right)^5(H\tau)^4+\frac{1}{2}\left(\frac{\sigma}{\mu}\right)^3 \right]e^{\left(\frac{\mu}{\sigma}\right)^2\left[-\frac{(H\tau)^2}{4}+\frac{5(H\tau)^4}{48}-\frac{61(H\tau)^6}{1440} \right]}\simeq -\frac{5}{2}\frac{H^6}{4\pi^2}.
\eeq
The sum of contributions in this limit is the standard result $\mathcal{N}=9H^5/4\pi^2$. This is confirmed using the exact mode solution \eqref{eq:SRdSmodesolution} in \eqref{eq:noisecorr2}, the noise amplitude of which is illustrated in Fig. \ref{fig_Erfc_IntegratedNoiseVSsigmaExact}. From this we see that the long-wavelength approximate mode solution \eqref{eq:IRappmode} used in \eqref{eq_ErfcNoiseIRapproxAnalytical} is accurate for cutoff scales with $\mu\lesssim 0.1$ for small values of $\sigma$.

The dependence on $\sigma$ is quite non-trivial. In particular, both positive and negative correlation regions arise as the ratio $\mu/\sigma$ increases, as can be seen in Fig.~\ref{fig:erfctaudep}. For illustration we have also included the Gaussian window function result at the same value of $\sigma$. 
While a Gaussian distribution for small $\sigma$ may be said to be well-approximated by a strict delta-function and a noise amplitude, a substantial amount of
information is lost with such a replacement when the distribution is both negative and
positive.

\paragraph{Splitting up the noise contributions.}
\begin{figure}[ht]
\centering
\begin{minipage}{0.45\textwidth}
\begin{tikzpicture}
    \begin{axis}[width=1.1\textwidth,height=1.1\textwidth-2cm,
    xmin=-4,
    xmax=4,
    ymin=-15,
    ymax=35,
    scaled ticks=false,
    xlabel = {$H\tau$},
    ylabel = {$\times\frac{4\pi^2}{H^6}$},
    ylabel shift = -0.1cm,
    ytick = {-10,0,10,20,30},
    scaled ticks=false,
    legend cell align={left},
    ]
    \addplot[blue] table [col sep=comma] {figures/fieldMomentumNoiseCorrs/XiphiXiphiErfcData.csv};
    \addlegendentry{$9H^2\av{{\xi}_\phi{\xi}_\phi}$},
    \addplot[red] table [col sep=comma] {figures/fieldMomentumNoiseCorrs/dotXidotXiErfcData.csv};
    \addlegendentry{$\av{\dot{\xi}_\phi\dot{\xi}_\phi}$}
    \addplot[green] table [col sep=comma] {figures/fieldMomentumNoiseCorrs/XipiXipiErfcData.csv};
    \addlegendentry{$\av{\xi_\pi\xi_\pi}$}
    \addplot[black,dotted] table [col sep=comma] {figures/fieldMomentumNoiseCorrs/CrossXiphidotXiphiErfcData.csv};
    \addlegendentry{$3H\sum\av{\dot{\xi}_\phi\xi_\phi}$}
    \addplot[black,dashed] table [col sep=comma] {figures/fieldMomentumNoiseCorrs/CrossXipiXiphiErfcData.csv};
    \addlegendentry{$3H\sum\av{{\xi}_\pi\xi_\phi}$}
    \addplot[gray] table [col sep=comma] {figures/fieldMomentumNoiseCorrs/CrossXipidotXiphiErfcData.csv};
    \addlegendentry{$\sum\av{\dot{\xi}_\phi\xi_\pi}$}
    \end{axis}
\end{tikzpicture}
\end{minipage}\hfill
\begin{minipage}{0.45\textwidth}
\hspace{-1cm}
\begin{tikzpicture}
    \begin{axis}[width=1.1\textwidth,height=1.1\textwidth-2cm,
    xmin=0,
    xmax=0.048,
    ymin=-1,
    ymax=10,
    scaled ticks=false,
    xlabel = {$\epsilon_M$},
    ylabel = {$\times\frac{4\pi^2}{H^5}$},
    ylabel shift = -0.1cm,
    scaled ticks=false,
    legend cell align={left},
    ytick = {0,3,6,9},
    xtick = {0.01,0.02,0.03,0.04},
    xticklabel style={
    /pgf/number format/precision=2,
    /pgf/number format/fixed,
    /pgf/number format/fixed zerofill,
    },
    ]
    \addplot[blue] table [col sep=comma] {figures/massdepDecomposedCorrsErfc/NoiseAmpXiphiXiphiErfcVSmassData.csv};
    \addlegendentry{$9H^2\mathcal{N}_{\xi_\phi\xi_\phi}$},
    \addplot[red] table [col sep=comma] {figures/massdepDecomposedCorrsErfc/NoiseAmpdotXidotXiErfcVSmassData.csv};
    \addlegendentry{$\mathcal{N}_{\dot{\xi}_\phi\dot{\xi}_\phi}$}
    \addplot[green] table [col sep=comma] {figures/massdepDecomposedCorrsErfc/NoiseAmpXipiXipiErfcVSmassData.csv};
    \addlegendentry{$\mathcal{N}_{\xi_\pi\xi_\pi}$}
    \addplot[black,dotted] table [col sep=comma] {figures/massdepDecomposedCorrsErfc/CrossCombAmpXiphidotXiphiErfcVSmassData.csv};
    \addlegendentry{$3H\sum\mathcal{N}_{\dot{\xi}_\phi\xi_\phi}$}
    \addplot[black,dashed] table [col sep=comma] {figures/massdepDecomposedCorrsErfc/CrossCombAmpXipiXiphiErfcVSmassData.csv};
    \addlegendentry{$3H\sum\mathcal{N}_{{\xi}_\pi\xi_\phi}$}
    \addplot[gray] table [col sep=comma] {figures/massdepDecomposedCorrsErfc/CrossCombAmpXipidotXiphiErfcVSmassData.csv};
    \addlegendentry{$\sum\mathcal{N}_{\dot{\xi}_\phi\xi_\pi}$}
    \end{axis}
\end{tikzpicture}
\end{minipage}
\caption{Left panel: Noise correlators for the Erfc window function using $\epsilon_M=0.01$, $r/R=0.01$ and values $\sigma=0.005$, $\mu=0.03$. As the ratio $\mu/\sigma$ increases the visible curves sharpen around $H\tau=0$ and the visible peaks grow towards $\pm\infty$. Right panel: The corresponding noise amplitude contributions as a function of the mass parameter $\epsilon_M$. }
\label{fig:fieldMomentumErfcNoises}
\end{figure}
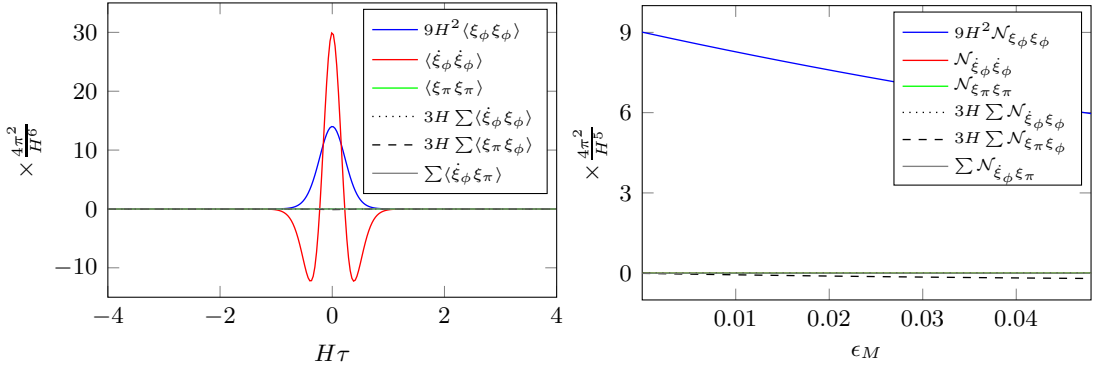

It is possible to identify the origin of both negative and positive correlations, and the dominant contribution to the noise amplitude, by studying the decomposed noise correlations $\xi_\phi,\xi_\pi$ of \eqref{eq:decomposedNoises}.
In particular, the visible negative correlation regions originate from the $\av{\dot{\xi}_\phi\dot{\xi}_\phi}$ correlator with $\xi_\phi$ given by \eqref{schemNoiseExprs}. In  \eqref{eq:dotXinoise} only the first term contributes in the massless limit, and so
\beq
    \av{\dot{\xi}_\phi(x)\dot{\xi}_\phi(x')}&\simeq \frac{H^6}{8\pi^3}e^{\frac{\mu^2}{\sigma^2}\left(\frac{\sech H\tau}{\sech^2\frac{H\tau}{2}}-1\right)}\bigg[\frac{1}{\sigma^6}\bigg(\tilde{\mathcal{T}}_{-5}
    -\frac{2\mu}{\sech\frac{H\tau}{2}}\tilde{\mathcal{T}}_{-4}+\mu^2\mathcal{T}_{-3}\bigg)\\
    &\qquad -\frac{2}{\sigma^4}\bigg(\frac{1}{\sech H\tau }\tilde{\mathcal{T}}_{-3} 
    -\frac{\mu}{\sech\frac{H\tau}{2}}\tilde{\mathcal{T}}_{-2} \bigg)+\frac{\tilde{\mathcal{T}}_{-1}}{\sigma^2} \bigg].
\eeq
Given the parametric choices made above, the integral of this function over $\tau$ vanishes, and so does not contribute to the noise amplitude in our definition. Nevertheless, the function itself is both positive and negative.
Similarly, for the cross terms, 
\beq
    \av{\dot{\xi}_\phi(x){\xi}_\phi(x')}+\av{{\xi}_\phi(x)\dot{\xi}_\phi(x')} &\simeq  \frac{H^5}{8\pi^3}e^{\frac{\mu^2}{\sigma^2}\left(\frac{\sech H\tau}{\sech^2\frac{H\tau}{2}}-1\right)}\\
    &\quad\times\bigg[\frac{2}{\sigma^4}\bigg(\frac{1}{\sech H\tau }\tilde{\mathcal{T}}_{-3} 
    -\frac{\mu}{\sech\frac{H\tau}{2}}\tilde{\mathcal{T}}_{-2} \bigg)-\frac{2}{\sigma^2}\tilde{\mathcal{T}}_{-1} \bigg],
\eeq
which vanish by \eqref{eq:tildeTapprox}.
The decomposed noise correlators, including also $\xi_\pi$, are all illustrated in Fig. \ref{fig:fieldMomentumErfcNoises}. 
From this we can conclude that although the resulting full noise correlator is rather complicated, the main contribution to the noise amplitude comes from the component
\beq 
    \av{\xi_\phi\xi_\phi} &= -\frac{iH^4}{16\pi^3}\frac{e^{M_1-\left(\frac{\mu}{\sigma}\right)^2}}{\sigma^2}\frac{R}{r}\mathcal{T}_{-(1+2\epsilon_M)},
    \label{eq:erfcxiphixiphiCorr}
\eeq
whose behaviour when expanded near $\tau=0$ is well-described by a Gaussian-like, positive function 
\beq 
    \av{\xi_\phi\xi_\phi} &\simeq  \frac{H^4}{4\pi^2}\frac{\mu^{2\epsilon_M}}{2\sqrt{\pi}}{e^{-\frac{1}{4}\left(\frac{\mu}{\sigma}\right)^2(H\tau)^2}}
    \frac{\mu}{\sigma}
    \bigg[1+\frac{\epsilon_M}{2}\Big(\frac{\sigma}{\mu}\Big)^2\\
    &\qquad-\frac{1}{8}\Big(5+\frac{3}{2}\Big(4+\Big(\frac{\sigma}{\mu}\Big)^2\Big)\epsilon_M\Big)(H\tau)^2
    +\mathcal{O}\Big(\Big(\frac{\sigma}{\mu}\Big)^4\Big)\Big].
\eeq
For small values of $\sigma/\mu$, this is approximated rather well by a delta function. The noise amplitude in this case is then given by
\beq 
    \mathcal{N}_{\xi_\phi\xi_\phi}  &\simeq \frac{H^3}{4\pi^2}{\mu^{2\epsilon_M}}\Big[1-\Big(\frac{\sigma}{\mu}\Big)^2\Big(\frac{5}{4}+\epsilon_M\Big)
    +\mathcal{O}\Big(\Big(\frac{\sigma}{\mu}\Big)^4\Big)\Big]\equiv \frac{H^3}{4\pi^2}f(\epsilon_M,\mu,\sigma).
    \label{erfc_f1}
\eeq
We now have an explicit example of the correction function $f$ advertised in section \ref{sec:introduction}. It involves the mass through $\epsilon_M$ as well as the parameters of the window function in the combination $\sigma/\mu$ with an overall factor of $\mu^{2\epsilon_M}$.

\begin{figure}
    \centering
    \begin{subfigure}[t]{0.45\textwidth}
    \begin{tikzpicture}
    \begin{axis}[width=1.1\textwidth,height=1.1\textwidth-2cm,
    scaled ticks=false,
    xmin=0,
    xmax=0.05,
    ymin=4.6,
    ymax=9.3,
    ytick = {5,6,7,8,9},
    xlabel = { $\epsilon_M$},
    ylabel = {$\frac{4\pi^2}{H^5}\mathcal{N}$},
    ylabel shift = -0.1cm,
    xticklabel style={
    /pgf/number format/precision=2,
    /pgf/number format/fixed,
    /pgf/number format/fixed zerofill,
    },
    scaled ticks=false,
    legend cell align={left},
    legend pos=south west,
    ]
    \addplot[blue] table [col sep=comma] {figures/massdep/GaussianNoiseAmplVSmassData.csv};
    \addlegendentry{Gaussian}
    \addplot[purple] table [col sep=comma] {figures/massdep/ErfNoiseAmplVSmassMu0025Data.csv};
    \addlegendentry{Erfc $\mu=0.025$}
    \addplot[red] table [col sep=comma] {figures/massdep/ErfNoiseAmplVSmassMu0030Data.csv};
    \addlegendentry{Erfc $\mu=0.030$}
    \addplot[orange] table [col sep=comma] {figures/massdep/ErfNoiseAmplVSmassMu0050Data.csv};
    \addlegendentry{Erfc $\mu=0.050$}
    \addplot[black,dashed] table [col sep=comma] {figures/massdep/prefacMuEpsilonDat.csv};
    \addplot[black,dotted] table [col sep=comma] {figures/massdep/prefacsigmaEpsilonDat.csv};
    \end{axis}
    \end{tikzpicture}
    \end{subfigure}
    \hfill
    \begin{subfigure}[t]{0.45\textwidth}
    \hspace{-1cm}
    \begin{tikzpicture}
    \begin{axis}[width=1.1\textwidth,height=1.1\textwidth-2cm,
    scaled ticks=false,
    xmin=0.02,
    xmax=0.1,
    ymin=8.3,
    ymax=9.1,
    ytick = {8.4,8.6,8.8,9.0},
    xtick = {0.02,0.04,0.06,0.08,0.1},
    xlabel = { $\mu$},
    ylabel shift = -0.1cm,
    yticklabel style={
    /pgf/number format/precision=1,
    /pgf/number format/fixed,
    /pgf/number format/fixed zerofill,
    },
    xticklabel style={
    /pgf/number format/precision=2,
    /pgf/number format/fixed,
    /pgf/number format/fixed zerofill,
    },
    ylabel = {$\frac{4\pi^2}{H^5}\mathcal{N}$},
    legend cell align={left},
    legend pos=south east,
    ]
    \addplot[purple] table [col sep=comma] {figures/mudepErfc/ErfcAmplVSmuZeroMass.csv};
    \addlegendentry{$\epsilon_M=0$}
    \addplot[red] table [col sep=comma] {figures/mudepErfc/ErfcAmplVSmuMass0dot001.csv};
    \addlegendentry{$\epsilon_M=0.001$}
    \addplot[orange] table [col sep=comma] {figures/mudepErfc/ErfcAmplVSmuMass0dot002.csv};
    \addlegendentry{$\epsilon_M=0.002$}
    \addplot[yellow] table [col sep=comma] {figures/mudepErfc/ErfcAmplVSmuMass0dot005.csv};
    \addlegendentry{$\epsilon_M=0.005$}
    \end{axis}
    \end{tikzpicture}
    \end{subfigure}
    \caption{Left panel: Noise amplitude for the Gaussian and Erfc window functions using exact mode functions as a function of $\epsilon_M$ with $\sigma=0.005$ and $r/R=0.01$. Dashed line is $9\mu^{2\epsilon_M}$ with $\mu=0.05$ and dotted line is $9\sigma^{2\epsilon_M}$. Right panel: Cutoff dependence of the noise amplitude with the Erfc window function for different choices of $\epsilon_M$.}
    \label{fig:massdeperfc}
\end{figure}
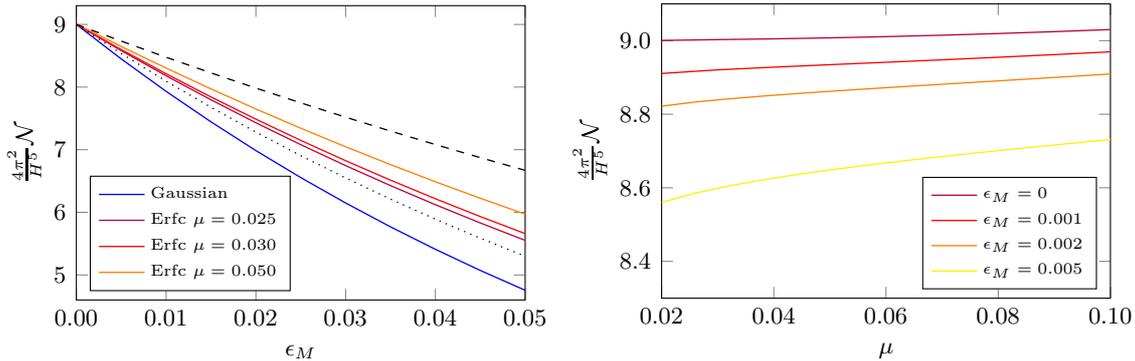

In the massless case, we get 
\beq
    f = 1-\frac{5}{4}\Big(\frac{\sigma}{\mu}\Big)^2, \quad\epsilon_M=0.
\eeq
The standard result is then recovered upon taking the step-function limit $\sigma\rightarrow0$. With a massive field, even in that limit we obtain
\beq
    f = \mu^{2\epsilon_M},\quad \sigma\rightarrow0,
\eeq
similar to the Gaussian window function, where the parameter $\sigma$ also plays the role of $\mu$.
For small values of $\mu$ and $\sigma$, a nonzero mass $\epsilon_M\neq 0$ has the effect of decreasing the noise amplitude, as can be seen in Fig. \ref{fig:massdeperfc}. The dependence on mass occurs largely, but not entirely, through the prefactor $\mu^{2\epsilon_M}$ (or $\sigma^{2\epsilon_M}$ in the Gaussian case). Any amplitude smaller than $9H^5/4\pi^2$ can be achieved by making a different choice of parameters.

\subsection{Alternative smoothened step window functions}
\label{sec:otherwindows}
To illustrate that these results are fairly generic, we
will also consider two other sigmoid functions, 
\begin{align}
    W_\text{tanh}(k,t) &= \frac{1}{2}\Big(1+\tanh\Big[\frac{1}{\sigma}\Big(\frac{k}{a_tH}-\mu\Big)\Big]\Big),\\
    W_\text{arct}(k,t) &= \frac{1}{2}\Big(1+\frac{2}{\pi}\arctan\Big[\frac{\pi}{2\sigma}\Big({\frac{k}{a_tH}-\mu}\Big)\Big]\Big),
\end{align}
for which
\begin{align}
    \tilde{\mathcal{Q}}_\text{tanh}(k,t) &= -H^2\frac{k}{\sigma a_tH}\sech^2\Big[\frac{1}{\sigma}\Big(\frac{k}{a_tH}-\mu\Big)\Big]\Big(\frac{k}{\sigma a_tH}\tanh\Big[\frac{1}{\sigma}\Big(\frac{k}{a_tH}-\mu\Big)\Big] +1-\epsilon_M\Big),
\end{align}
and
\begin{align}
    \tilde{\mathcal{Q}}_\text{arct}(k,t) &= -H^2\frac{\frac{k}{\sigma a_tH}}{1+(\tfrac{\pi}{2\sigma})^2(\frac{k}{a_tH}-\mu)^2}\left[\frac{(\frac{\pi}{2\sigma})^2\frac{k}{ a_tH}(\frac{k}{a_tH}-\mu)}{1+(\tfrac{\pi}{2\sigma})^2(\frac{k}{a_tH}-\mu)^2}+1-\epsilon_M \right].
\end{align}
Restricting ourselves to comparing numerical results, the noise amplitude $\mathcal{N}$ for all window functions is shown in Fig. \ref{fig_ExactintegratedNoiseAllFunctionsVSsigma}. We see that in the limit $\sigma\rightarrow0$, they all converge to the standard result, but for finite $\sigma$, they diverge from each other. The convergence for the Arctan window function to the value $9H^5/4\pi^2$ is particularly slow, and only agrees in the limit, whereas the other window functions start converging for finite values of $\sigma\simeq 0.04$. This qualitative behaviour persists for other values of $\mu$ and $\epsilon_M$ (see Fig. \ref{fig:massdepWindows}, right panel), and is likely a reflection of the fact that the Arctan window requires smaller values of $\sigma$ to reach the step function shape compared to the other window functions.
As for the Erfc window function, the noise correlations using the other sigmoid window functions also develop negative correlation regions for values $\mu/\sigma \gtrsim 3$, as illustrated in the left panel of Fig \ref{fig:massdepWindows}.
\begin{figure}[t]
\centering
\begin{minipage}{0.45\textwidth}
\hspace{-1cm}
\begin{tikzpicture}
    \begin{axis}[width=1.1\textwidth,height=1.1\textwidth-2cm,
    scaled ticks=false,
    xmin=0.001,
    xmax=0.1,
    ymin=5,
    ymax=10,
    ytick = {5,6,7,8,9,10},
    ylabel shift = -0.1cm,
    xticklabel style={
    /pgf/number format/precision=2,
    /pgf/number format/fixed,
    /pgf/number format/fixed zerofill,
    },
    xlabel = {$\sigma$},
    ylabel = {$\frac{4\pi^2}{H^5}\mathcal{N}$},
    legend cell align={left},
    scaled ticks=false,
    legend pos=south west,
    ]
    \addplot[blue] table [col sep=comma] {figures/sigmadepcollected/GaussianNoiseNormVSsigma2.csv};
    \addlegendentry{Gaussian}
    \addplot[red] table [col sep=comma] {figures/sigmadepcollected/ErfcNoiseNormVSsigma.csv};
    \addlegendentry{Erfc}
    \addplot[green] table [col sep=comma] {figures/sigmadepcollected/TanhNoiseNormVSsigma.csv};
    \addlegendentry{Tanh}
    \addplot[orange] table [col sep=comma] {figures/sigmadepcollected/ArctanNoiseNormVSsigma.csv};
    \addlegendentry{Arctan}
    \end{axis}
    \end{tikzpicture}
    \end{minipage}
\begin{minipage}{0.45\textwidth}
\hspace{-0.5cm}
\begin{tikzpicture}
    \begin{axis}[width=1.1\textwidth,height=1.1\textwidth-2cm,
    scaled ticks=false,
    xmin=0.001,
    xmax=0.1,
    ymin=5,
    ymax=10,
    ytick = {5,6,7,8,9,10},
    ylabel shift = -0.1cm,
    xticklabel style={
    /pgf/number format/precision=2,
    /pgf/number format/fixed,
    /pgf/number format/fixed zerofill,
    },
    xlabel = {$\sigma$},
    ylabel = {$\frac{4\pi^2}{H^5}\mathcal{N}$},
    scaled ticks=false,
    legend cell align={left},
    legend pos=south west,
    ]
    \addplot[blue] table [col sep=comma] {figures/sigmadepcollected/GaussianNoiseAmplVSsigmaDataSmallmass.csv};
    \addlegendentry{Gaussian}
    \addplot[red] table [col sep=comma] {figures/sigmadepcollected/ErfNoiseAmplVSsigmaDataSmallmass.csv};
    \addlegendentry{Erfc}
    \addplot[green] table [col sep=comma] {figures/sigmadepcollected/TanhNoiseAmplVSsigmaDataSmallmass.csv};
    \addlegendentry{Tanh}
    \addplot[orange] table [col sep=comma] {figures/sigmadepcollected/ArctanNoiseAmplVSsigmaDataSmallmass.csv};
    \addlegendentry{Arctan}
    \end{axis}
    \end{tikzpicture}
\end{minipage}    
    \caption{Noise amplitude for different window functions using the exact propagator \eqref{eq:SRdSmodesolution} as a function of $\sigma$ with masses $\epsilon_M=0$ (left panel) and $\epsilon_M=0.002$ (right panel) 
    using $\mu=0.1$ and $r/R=0.01$.}
    \label{fig_ExactintegratedNoiseAllFunctionsVSsigma}
\end{figure}
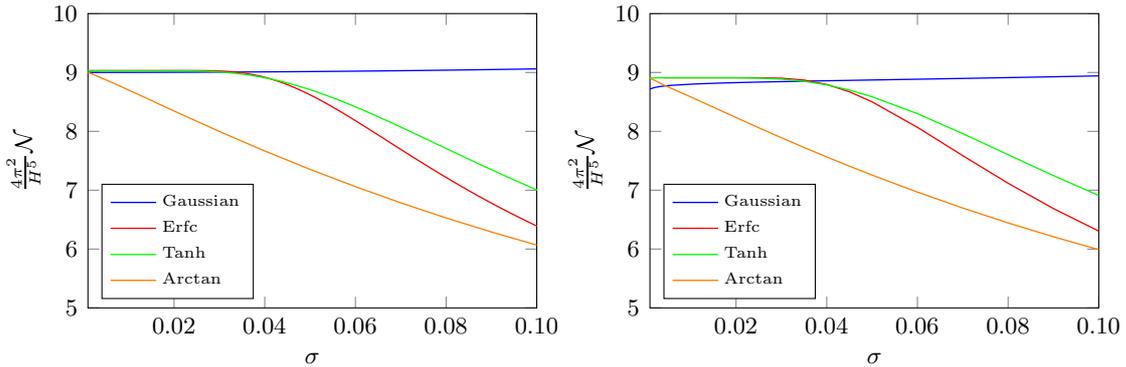
\begin{figure}[ht]
    \centering
    \begin{subfigure}[t]{0.45\textwidth}
    \begin{tikzpicture}
    \begin{axis}[width=1.1\textwidth,height=1.1\textwidth-2cm,
    scaled ticks=false,
    xmin=-5,
    xmax=5,
    ymin=-20,
    ymax=80,
    ytick = {-20,0,20,40,60},
    ylabel shift = -0.1cm,
    xlabel = { $H\tau$},
    ylabel = {$\frac{4\pi^2}{H^6}\av{\xi(x)\xi(x')}$},
    legend pos=north east,
    legend cell align={left},
    ]
    \addplot[blue] table [col sep=comma] {figures/timesepdep/GnoiseSmallmassData.csv};
    \addlegendentry{Gaussian}
    \addplot[red] table [col sep=comma] {figures/timesepdep/EnoiseSmallmassData.csv};
    \addlegendentry{Erfc}
    \addplot[green] table [col sep=comma] {figures/timesepdep/TnoiseSmallmassData.csv};
    \addlegendentry{Tanh}
    \addplot[orange] table [col sep=comma] {figures/timesepdep/AnoiseSmallmassData.csv};
    \addlegendentry{Arctan}
    \end{axis}
    \end{tikzpicture}
    \end{subfigure}
    \hfill
    \begin{subfigure}[t]{0.45\textwidth}
    \hspace{-1cm}
    \begin{tikzpicture}
    \begin{axis}[width=1.1\textwidth,height=1.1\textwidth-2cm,
    scaled ticks=false,
    xmin=0,
    xmax=0.05,
    ymin=4.5,
    ymax=9.5,
    ytick = {5,6,7,8,9},
    xlabel = { $\epsilon_M$},
    ylabel = {$\frac{4\pi^2}{H^5}\mathcal{N}$},
    ylabel shift = -0.1cm,
    xticklabel style={
    /pgf/number format/precision=2,
    /pgf/number format/fixed,
    /pgf/number format/fixed zerofill,
    },
    legend pos=north east,
    legend cell align={left},
    ]
    \addplot[blue] table [col sep=comma] {figures/massdep/GaussianNoiseAmplVSmassData.csv};
    \addlegendentry{Gaussian}
    \addplot[red] table [col sep=comma] {figures/massdep/ErfNoiseAmplVSmassData2.csv};
    \addlegendentry{Erfc}
    \addplot[green,dashed] table [col sep=comma] {figures/massdep/TanhNoiseAmplVSmassData.csv};
    \addlegendentry{Tanh}
    \addplot[orange] table [col sep=comma] {figures/massdep/ArctanNoiseAmplVSmassData.csv};
    \addlegendentry{Arctan}
    \end{axis}
    \end{tikzpicture}
    \end{subfigure}
    \caption{Left panel: The noise correlations for different window functions as a function of time separation with $\epsilon_M=0.01$, $r/R=0.01$, $\sigma=0.005$ and in addition $\mu=0.03$ for the sigmoids. Right panel: The corresponding noise amplitudes as a function of $\epsilon_M$. For small $\sigma$ the Erfc and Tanh windows integrate to very similar values. Smaller values of $\sigma$ are required for the Arctan window function to recover the higher noise amplitudes of the other smooth step functions.}
    \label{fig:massdepWindows}
\end{figure}
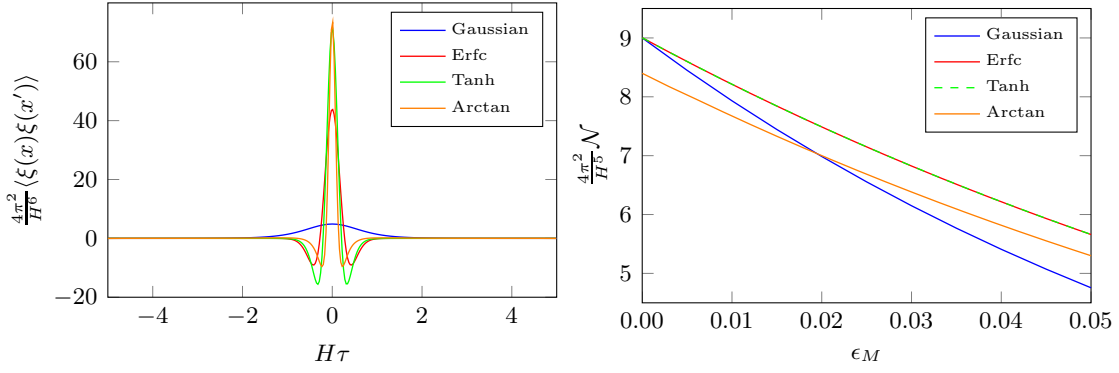

\subsection{Away from dS: leading order in slow-roll}
\label{sec:slowroll}
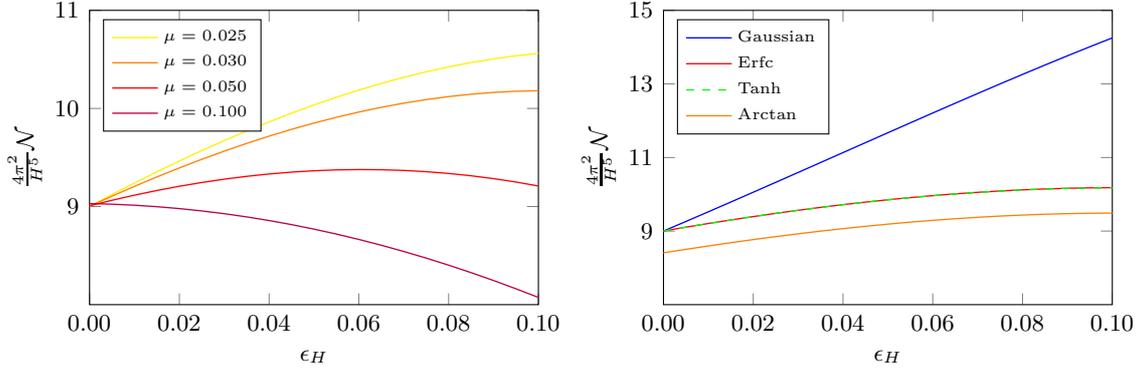
\begin{figure}
    \centering
    \begin{subfigure}[b]{0.45\textwidth}
    \hspace{-0.7cm}
    \begin{tikzpicture}
    \begin{axis}[width=1.1\textwidth,height=1.1\textwidth-2cm,
    scaled ticks=false,
    xmin=0,
    xmax=0.1,
    ymin=8,
    ymax=11,
    ytick = {9,10,11},
    xlabel = { $\epsilon_H$},
    ylabel = {$\frac{4\pi^2}{H^5}\mathcal{N}$},
    ylabel shift = -0.1cm,
    xticklabel style={
    /pgf/number format/precision=2,
    /pgf/number format/fixed,
    /pgf/number format/fixed zerofill,
    },
    legend pos=north west,
    legend cell align={left},
    ]
    \addplot[yellow] table [col sep=comma] {figures/epsilonHdep/ErfcAmpVSepsilonHdatMu=0_025.csv};
    \addlegendentry{$\mu=0.025$}
    \addplot[orange] table [col sep=comma] {figures/epsilonHdep/ErfcAmpVSepsilonHdatMu=0_03.csv};
    \addlegendentry{$\mu=0.030$}
    \addplot[red] table [col sep=comma] {figures/epsilonHdep/ErfcAmpVSepsilonHdatMu=0_05.csv};
    \addlegendentry{$\mu=0.050$}
    \addplot[purple] table [col sep=comma] {figures/epsilonHdep/ErfcAmpVSepsilonHdatMu=0_1.csv};
    \addlegendentry{$\mu=0.100$}
    \end{axis}
    \end{tikzpicture}
    \end{subfigure}
    \begin{subfigure}[b]{0.45\textwidth}
    \begin{tikzpicture}
    \begin{axis}[width=1.1\textwidth,height=1.1\textwidth-2cm,
    scaled ticks=false,
    xmin=0,
    xmax=0.1,
    ymin=7,
    ymax=15,
    ytick = {9,11,13,15},
    ylabel shift = -0.1cm,
    xlabel = { $\epsilon_H$},
    ylabel = {$\frac{4\pi^2}{H^5}\mathcal{N}$},
    xticklabel style={
    /pgf/number format/precision=2,
    /pgf/number format/fixed,
    /pgf/number format/fixed zerofill,
    },
    legend pos=north west,
    legend cell align={left},
    ]
    \addplot[blue] table [col sep=comma] {figures/epsilonHdep/GaussianNoiseAmplVSepsilonHdata.csv};
    \addlegendentry{Gaussian}
    \addplot[red] table [col sep=comma] {figures/epsilonHdep/ErfNoiseAmplVSepsilonHdata.csv};
    \addlegendentry{Erfc}
    \addplot[green,dashed] table [col sep=comma] {figures/epsilonHdep/TanhNoiseAmplVSepsilonHdata.csv};
    \addlegendentry{Tanh}
    \addplot[orange] table [col sep=comma] {figures/epsilonHdep/ArctanNoiseAmplVSepsilonHdata.csv};
    \addlegendentry{Arctan}
    \end{axis}
    \end{tikzpicture}
    \end{subfigure}
    \caption{Left panel: Noise amplitude with the Erfc window function as a function of the slow-roll parameter $\epsilon_H$ for different values of $\mu$. Right panel: For all window functions for a cutoff $\mu=0.03$. Here $\epsilon_M=0$, $\sigma=0.005$ and $r/R=0.01$. The Arctan window function requires even smaller values of $\sigma$ to reach the asymptotic value $9H^5/4\pi^2$ in the $\epsilon_H\rightarrow 0$ limit. }
    \label{fig:slow-rolldepWindows}
\end{figure}
As a final generalisation, we consider the dependence of the slow-roll parameter $\epsilon_H$. Using that the above window functions depend on $k$ through the combination ${W}(t,k)\equiv {W}(k/\sigma a(t)H(t))$, it is convenient to make the replacement, 
\begin{align}
    \Dot{{W}}_t &=-Hz(1-\epsilon_H)\partial_{z}{W}_t(z),\qquad z\equiv \frac{k}{\sigma a(t)H(t)},\\
    \Ddot{{W}}_t &\simeq H^2(1-\epsilon_H)z\partial_z{W}(z)+H^2(1-2\epsilon_H)z^2\partial_z^2{W}(z),
\end{align}
where the last equality holds to first order in slow-roll. The function $\tilde{\mathcal{Q}}_t$ in the correlator \eqref{eq:noisecorr2} is then
\beq
\begin{aligned}
    -\tilde{\mathcal{Q}}_t&=\Ddot{{W}}_t+(3+2q_\nu)H(1-\epsilon_H)\Dot{{W}}_t\simeq H^2(1-2\epsilon_H)\left[z^2\partial_z^2-2(1+q_\nu)z\partial_z\right]{W}(z).
\end{aligned}
\eeq

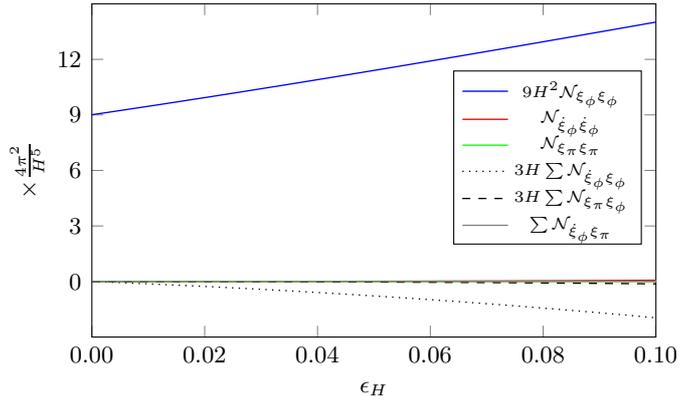
\begin{figure}[htb]
\centering
\begin{tikzpicture}
    \begin{axis}[width=9cm,height=6cm,
    xmin=0,
    xmax=0.1,
    ymin=-3,
    ymax=15,
    scaled ticks=false,
    ylabel shift = -0.1cm,
    xtick={0.00,0.02,0.04,0.06,0.08,0.10},
    xlabel = {$\epsilon_H$},
    ylabel = {$\times\frac{4\pi^2}{H^5}$},
    scaled ticks=false,
    legend style={at={(.975,.8)}},
    ytick = {0,3,6,9,12},
    xticklabel style={
    /pgf/number format/precision=2,
    /pgf/number format/fixed,
    /pgf/number format/fixed zerofill,
    },
    ]
    \addplot[blue] table [col sep=comma] {figures/epsilonHdep/NoiseAmpXiphiXiphiErfcVSepsilonHData.csv};
    \addlegendentry{$9H^2\mathcal{N}_{\xi_\phi\xi_\phi}$},
    \addplot[red] table [col sep=comma] {figures/epsilonHdep/NoiseAmpdotXidotXiErfcVSepsilonHData.csv};
    \addlegendentry{$\mathcal{N}_{\dot{\xi}_\phi\dot{\xi}_\phi}$}
    \addplot[green] table [col sep=comma] {figures/epsilonHdep/NoiseAmpXipiXipiErfcVSepsilonHData.csv};
    \addlegendentry{$\mathcal{N}_{\xi_\pi\xi_\pi}$}
    \addplot[black,dotted] table [col sep=comma] {figures/epsilonHdep/CrossCombAmpXiphidotXiphiErfcVSepsilonHData.csv};
    \addlegendentry{$3H\sum\mathcal{N}_{\dot{\xi}_\phi\xi_\phi}$}
    \addplot[black,dashed] table [col sep=comma] {figures/epsilonHdep/CrossCombAmpXipiXiphiErfcVSepsilonHData.csv};
    \addlegendentry{$3H\sum\mathcal{N}_{{\xi}_\pi\xi_\phi}$}
    \addplot[gray] table [col sep=comma] {figures/epsilonHdep/CrossCombAmpXipidotXiphiErfcVSepsilonHData.csv};
    \addlegendentry{$\sum\mathcal{N}_{\dot{\xi}_\phi\xi_\pi}$}
    \end{axis}
\end{tikzpicture}
\caption{Slow-roll dependence of the noise amplitudes of the decomposed correlators \eqref{eq:decomposedNoises} with the Erfc window function, in the massless case, for $r/R=0.01$, $\sigma=0.005$ and $\mu=0.03$.}
\label{fig:epsilonHdepDecomposedNoisesErfc}
\end{figure}

The slow-roll parameter $\epsilon_H$ appears both in the exact mode functions through the index $\nu$, and through the derivative operators $\tilde{\mathcal{Q}}_t$. The rest of the computation proceeds in the same way as before. For instance for the Erfc window with the long-wavelength approximate solution \eqref{eq:IRappmode}, the normalisation becomes
\beq 
    \mathcal{N}_{\xi_\phi\xi_\phi} &\simeq \frac{H^3}{4\pi^2}\frac{(1-\epsilon_H)^4}{1-\frac{1}{2}\epsilon_H}\bigg(\frac{\mu}{1-\epsilon_H}\bigg)^{2\epsilon_M-2\epsilon_H}\bigg[1-\Big(\frac{\sigma}{\mu}\Big)^2\Big(\frac{5}{4}+\epsilon_M-\epsilon_H\Big)
    +\mathcal{O}\Big(\Big(\frac{\sigma}{\mu}\Big)^4\Big) \bigg]\\
    &\equiv \frac{H^3}{4\pi^2}f(\epsilon_M,\epsilon_H,\mu,\sigma).
\eeq
This is a further generalisation of (\ref{erfc_f1}), where now also different values of $\epsilon_H$ may alter the noise normalisation. 

Inserting the exact mode solutions \eqref{eq:SRdSmodesolution}, we evaluate the noise numerically for the different window functions considered, and the resulting $\epsilon_H$-dependence is shown in Fig. \ref{fig:slow-rolldepWindows}. We see that, as for $\epsilon_M$, the dependence on the slow-roll parameter is also substantial, non-monotonic and dependent on the choice of both window function and the cutoff parameters. Away from dS, the contributions to the noise amplitude is spread to the other noise correlators in \eqref{eq:decomposedNoises}, as illustrated in Fig. \ref{fig:epsilonHdepDecomposedNoisesErfc} for the Erfc window function. With $\mu, \sigma$ small, the correlator $\av{\xi_\phi\xi_\phi}$ remains the leading contributor to the noise amplitude.

\section{Conclusions}
\label{sec:conclusion}

In this work, we have revisited the derivation of stochastic inflation from first-principles quantum field theory. 
We have investigated the conditions under which the standard Langevin dynamics for the super-horizon component of a scalar field may be recovered. The agreement is subject to an appropriate choice of window function separating the UV and IR modes, and applies only in certain limits, which we have identified. 
We find that one must pay close attention to the magnitude of the off-diagonal components of the influence functional kernel \eqref{eq:offdiag}, and compare these to the diagonal contribution that parametrise the stochastic noise. Only for a steep window function with a separation scale in the far super-horizon limit ($\sigma,\mu\ll 1$) is it consistent to neglect them.

The essential ingredient in achieving the conventional Langevin evolution is to note that noise correlations are exponentially suppressed both outside the spatial horizon $(aH)^{-1}$, and for large time separations ($H\tau>1$). Inside the spatial horizon, one may then attempt to represent the correlations by a completely local noise amplitude, obtained through matching the integrated power. We showed that this may be a valid prescription in some cases (e.g. with a Gaussian window function), but that for others, important information may be lost (e.g. with the Erfc window function). 

Our analysis confirms that in the limits $\epsilon_M, \epsilon_H\rightarrow 0$, and $\mu,\sigma\rightarrow 0$, the standard noise amplitude $9H^5/4\pi^2$ is recovered, at least asymptotically. This corresponds to an IR field which is essentially the $k\simeq 0$ homogeneous mode. However, as soon as these parameters take on finite values, the noise amplitude is modulated by a function $f$, which in principle can assume any value.
We provided a number of examples of what such a function $f$ could look like, for a selection of window functions. We also showed numerically, that the long-wavelength approximation to the UV mode functions only matches the exact mode-function results for very small $\mu$ and $\sigma$.

Because the stochastic formalism relies on a sequence of approximations and assumptions, it would be interesting to further map out its region of validity by comparing to other approaches. These include matching to perturbative and resummed analytic results \cite{Serreau:2011fu,Serreau:2013psa,Garbrecht:2013coa,Kamenshchik:2021tjh}, and numerical evaluation within (generalisations of) the target stochastic model \cite{Cable:2020dke}. One may also consider a full numerical implementation of the quantum dynamics, in a semi-dS background, similar to what was done in Ref. \cite{Tranberg:2008ae}.

To summarise, stochastic inflation as a formalism remains a very elegant and powerful way of describing the dynamics of cosmological scalar fields at very long wavelengths. For it also to describe the dynamics of super-horizon modes in general ($k\neq 0$), the properties of the window function, the field mass and slow-roll parameters come into play. In particular, they enter in the explicit computation of the noise amplitude, which normalises observables, including the dynamical mass generated.

\bibliographystyle{elsarticle-num} 
\end{document}